\documentclass{emulateapj} %[12pt,preprint]{aastex}

\usepackage {epsf}
\usepackage{epstopdf}
\usepackage {lscape, graphicx}

\gdef\1054{MS\,1054--03}

\def\simgeq{{\raise.0ex\hbox{$\mathchar"013E$}\mkern-14mu\lower1.2ex\hbox{$\mathchar"0218$}}} 

\begin {document}

%\title {Recipes for Deriving Star Formation Rates of Galaxies out to $z \sim 3$}
\title {On Star Formation Rates and Star Formation Histories of Galaxies out to $z \sim 3$}
%\title {\small Short Title: SFRs and SFHs out to $z \sim 3$}

\author{Stijn Wuyts\altaffilmark{1}, Natascha M. F\"{o}rster
  Schreiber\altaffilmark{1}, Dieter Lutz\altaffilmark{1}, Raanan
  Nordon\altaffilmark{1}, Stefano Berta\altaffilmark{1}, Bruno
  Altieri\altaffilmark{2}, Paola
  Andreani\altaffilmark{3,4}, Herv\'{e}
  Aussel\altaffilmark{5}, Angel Bongiovanni\altaffilmark{6,7}, Jordi
  Cepa\altaffilmark{6,7}, Andrea Cimatti\altaffilmark{8}, Emanuele Daddi\altaffilmark{5}, David
  Elbaz\altaffilmark{5}, Reinhard
  Genzel\altaffilmark{1}, Anton M. Koekemoer\altaffilmark{9}, Benjamin Magnelli\altaffilmark{1}, Roberto
  Maiolino\altaffilmark{10}, Elizabeth J. McGrath\altaffilmark{11}, Ana P\'{e}rez Garc\'{i}a\altaffilmark{6,7}, Albrecht
  Poglitsch\altaffilmark{1}, Paola Popesso\altaffilmark{1}, Francesca Pozzi\altaffilmark{8}, Miguel Sanchez-Portal\altaffilmark{2},
  Eckhard Sturm\altaffilmark{1}, Linda Tacconi\altaffilmark{1}, Ivan Valtchanov\altaffilmark{2}}
\altaffiltext{1}{Max-Planck-Institut f\"{u}r extraterrestrische Physik, Giessenbachstrasse, D-85748 Garching, Germany}
\altaffiltext{2}{Herschel Science Centre, ESAC, Villanueva de la Ca\~{n}ada, 28691 Madrid, Spain}
\altaffiltext{3}{ESO, Karl-Schwarzschild-Str. 2, D-85748 Garching,  Germany}
\altaffiltext{4}{INAF, Osservatorio Astronomico di Trieste, via Tiepolo 11, 34143 Trieste, Italy}
\altaffiltext{5}{Laboratoire AIM, CEA/DSM-CNRS-Universit\'{e} Paris Diderot, IRFU/Service d'Astrophysique, B\^{a}t.709, CEA-Saclay, 91191 Gif-sur-Yvette Cedex, France}
\altaffiltext{6}{Instituto de Astrof\'{i}sica de Canarias, 38205 La Laguna, Spain}
\altaffiltext{7}{Departamento di Astronomia, Universit\`{a} di Bologna, via Ranzani 1, 40127 Bologna, Italy}
\altaffiltext{8}{Dipartimento di Astronomia, Universit\`{a} di Bologna, via Ranzani 1, 40127 Bologna, Italy}
\altaffiltext{9}{Space Telescope Science Institute, 3700 San Martin Drive, Baltimore, MD 21218, USA}
\altaffiltext{10}{INAF, Osservatorio Astronomico di Roma, via di Frascati 33, 00040 Monte Porzio Catone, Italy}
\altaffiltext{11}{University of California Observatories/Lick Observatory, University of California, Santa Cruz, CA 95064, USA}

\begin{abstract}
We compare multi-wavelength star formation rate (SFR) indicators out
  to $z \sim 3$ in the GOODS-South field.  Our analysis uniquely
  combines $U$-to-8$\mu$m photometry from FIREWORKS, MIPS 24
  $\mu$m and PACS 70, 100, and 160 $\mu$m photometry from the PEP
  survey, and $H\alpha$ spectroscopy from the SINS survey.  We
  describe a set of conversions that lead to a continuity across SFR
  indicators.  A luminosity-independent conversion from 24 $\mu$m to
  total infrared luminosity yields estimates of $L_{IR}$ that are in
  the median consistent with the $L_{IR}$ derived from PACS
  photometry, albeit with significant scatter.  Dust correction
  methods perform well at low to intermediate levels of star
  formation.  They fail to recover the total amount of star formation
  in systems with large $SFR_{IR} / SFR_{UV}$ ratios, typically
  occuring at the highest SFRs ($SFR_{UV+IR} \gtrsim 100\
  M_{\sun}/yr$) and redshifts ($z \gtrsim 2.5$) probed.  Finally, we
  confirm that $H\alpha$-based SFRs at $1.5 < z < 2.6$ are consistent
  with $SFR_{SED}$ and $SFR_{UV+IR}$ provided extra attenuation
  towards HII regions is taken into account
  ($A_{V,neb}=A_{V,continuum}/0.44$).  With the cross-calibrated SFR
  indicators in hand, we perform a consistency check on the star
  formation histories inferred from SED modeling.  We compare the
  observed SFR-M relations and mass functions at a range of redshifts
  to equivalents that are computed by evolving lower redshift galaxies
  backwards in time.  We find evidence for underestimated stellar ages
  when no stringent constraints on formation epoch are applied in SED
  modeling.  We demonstrate how resolved SED modeling, or
  alternatively deep UV data, may help to overcome this bias.  The age
  bias is most severe for galaxies with young stellar populations, and
  reduces towards older systems.  Finally, our analysis suggests that
  SFHs typically vary on timescales that are long (at least several
  100 Myr) compared to the galaxies' dynamical time.
\end{abstract}


\section {Introduction}
\label{intro.sec}

Determining when the stars were formed is one of the main goals in
galaxy evolution.  Two widely adopted approaches to this question are
to measure the assembled stellar mass at various lookback times (see,
for the local universe, Cole et al. 2001; Bell et al. 2003, at
intermediate redshifts Bundy et al. 2006; Borch et al. 2006; Pozzetti
et al. 2007; Vergani et al. 2008, and out to $z \sim 5$ Dickinson et
al. 2003; Fontana et al. 2003, 2004, 2006; Drory et al. 2004, 2005;
Elsner et al. 2008; P\'{e}rez-Gonz\'{a}lez et al. 2008; Marchesini et
al. 2009), or to quantify the rate of on-going star formation over
cosmic time (Madau et al. 1996; Lilly et al. 1996; Steidel et
al. 1999; Giavalisco et al. 2004; Schiminovich et al. 2005; Bouwens et
al. 2007, 2010).  While the latter should in principle integrate up to
the former, modulo stellar mass loss, it is currently heavily debated
whether or not the data, or rather the physical quantities estimated
from them, satisfy this continuity equation (Hopkins \& Beacom 2006;
Reddy \& Steidel 2009).

The relation between the instantaneous star formation rate (SFR) and
stellar mass of individual galaxies sheds light on how (in bursts or
gradually) and where (in galaxies of what mass) the star-forming
activity took place (Noeske et al. 2007; Elbaz et al. 2007; Daddi et
al. 2007a).  At a given mass, observations show a rapidly increasing
SFR with redshift (e.g., Lilly et al. 1996; Cowie et al. 1996; Bell et
al. 2005), whereas the models predict an increase in growth rate that
is significantly slower (Dav\'{e} 2008; Damen et al. 2009).

Both issues urge us to critically investigate the assumptions made in
translating fluxes and colors to an estimate of the SFR.  Especially
the presence of dust greatly impacts the interpretation of
multi-wavelength data.  The SFR indicators considered in this paper
come in two flavors: either the rate of unobscured and obscured star
formation is summed ($SFR_{UV+IR}$ in this paper), or the rate of
unobscured star formation is scaled up by a dust correction factor
($SFR_{\rm SED}$, $SFR_{UV, corr}$ and $SFR_{H\alpha}$ in this paper).
The former requires knowledge of the total infrared (IR) luminosity
($L_{IR}$) emitted in the rest-frame 8-1000 $\mu$m range by dust that
was heated by young stars, often extrapolated from the observed 24
$\mu$m photometry.  The latter requires knowledge of how the observed
colors break down in intrinsic colors of the stellar population and
dust reddening, which both are model dependent.

Here, we exploit the wealth of multi-wavelength data in the
GOODS-South field, including {\it HST}/ACS, {\it HST}/WFC3, VLT/ISAAC,
{\it Spitzer}/IRAC, {\it Spitzer}/MIPS 24 $\mu$m, and {\it
  Herschel}/PACS 70 - 160 $\mu$m imaging, as well as VLT/SINFONI
$H\alpha$ spectroscopy to calibrate SFR indicators of galaxies out to
$z \sim 3$.  This work lays the basis for Wuyts et al. (2011b), where
we analyze galaxy morphologies as a function of position in the
SFR-Mass diagram.  In order to carry out such a systematic study of
the entire galaxy population over a large range of masses and SFRs,
establishing a continuity across SFR indicators is essential.  This is
obviously of more general use as well.  Doing so, our analysis places
constraints on the distribution of dust within these galaxies, and on
their star formation histories (SFHs).  In addition, we exploit the
inferred SFRs and SFHs to test whether galaxy populations at different
lookback times satisfy a continuity equation.

We present a brief overview of the data used in Section\
\ref{sample.sec}.  Next, we make the case for a luminosity-independent
conversion to $L_{IR}$ in Section\ \ref{Lir.sec}.  SFRs derived by
summing a UV and IR contribution are contrasted to SED modeled SFRs
and dust-corrected UV SFRs in Section\
\ref{SFR_SED.sec}, and to $H\alpha$-derived SFRs in Section\
\ref{SFR_Ha.sec}.  Section\ \ref{consistency.sec} discusses the
consistency of inferred SFHs, and introduces resolved SED modeling as
a means to better characterize the age of the bulk of the stars in a
galaxy.  Finally, we summarize our results in Section\ \ref{discussion.sec}.

% Also investigate SFHs?  Increasing SFHs?

Throughout this paper, we quote magnitudes in the AB system, and
adopt the following cosmological parameters: $(\Omega _M, \Omega
_{\Lambda}, h) = (0.3, 0.7, 0.7)$.

\section {Sample and Observations}
\label{sample.sec}

\subsection {FIREWORKS $K_s$-selected Catalog}
\label{fireworks.sec}

The backbone of this work is the $K_s$-selected FIREWORKS catalog
sampling the $U$-to-8$\mu$m regime with 16 passbands ($K_{s, lim} =
24.3,\ 5\sigma$; Wuyts et al. 2008, hereafter W08).  Initially, we
focus on the 1333 galaxies in the redshift range $0 < z < 3$ that were
significantly detected ($> 3\sigma$) at 24$\mu$m.  Redshifts are
spectroscopic for 69\% of the sample below $z < 1.5$, and for 27\%
above $z > 1.5$.  For the remaining sources, we derived high-quality
photometric redshifts using EAZY (Brammer et al. 2008).  The median
and normalized median absolute deviation in $\Delta z / (1+z)$ are
$(-0.002;\ 0.025)$ at $z < 1.5$ and $(-0.003;\ 0.044)$ at $z > 1.5$.

\subsection {PEP GOODS-S Survey}
\label{PEP.sec}

The GOODS-South field has been observed with the Photodetector Array
Camera and Spectrometer (PACS, Poglitsch et al. 2010) onboard the {\it
  Herschel} Space Observatory as part of the PACS Extragalactic Survey
(PEP, Lutz et al. 2011).  With $5 \sigma$ depths for prior extraction of 1.8 mJy, 1.9 mJy
and 3.3 mJy at 70 $\mu$m, 100 $\mu$m, and 160 $\mu$m respectively, it
is the only PEP field to be imaged at 70 $\mu$m, and a factor 2.5 - 4.5
deeper than the other PEP blank fields at 100 $\mu$m and 160 $\mu$m.
These observed bands sample rest-frame wavelengths close to the peak
of IR emission of galaxies in our redshift range of interest.  For an
in depth description of the observations, data reduction and catalog
building, we defer to Berta et al. (2010) and Lutz et al. (2011).
Briefly, the PACS photometry used in this paper was obtained by
PSF-fitting using the positions of MIPS 24 $\mu$m detected sources
from Magnelli et al. (2009) as prior.  Adopting this
positional prior reduces the effects of confusion, and is justified by
the relative depths of the MIPS and PACS imaging.

\subsection {SINS GOODS-S Sample}
\label{SINS.sec}

The SINS survey (e.g., F\"{o}rster Schreiber et al. 2006, 2009; Genzel
et al. 2006, 2008) is a VLT/SINFONI GTO program targeting 80
high-redshift star-forming galaxies, providing integral-field emission
line maps and integrated fluxes without slit losses.  We focus on the
subset of 25 SINS galaxies located in the GOODS-South field with H$\alpha$ observations spanning
a redshift range $1.5 < z < 2.6$, and a range in stellar mass from
$5.1 \times 10^9 M_{\sun}$ to $1.4 \times 10^{11} M_{\sun}$.  Their
location in the SFR versus mass diagram is representative for that of
the underlying population of star-forming galaxies at similar
redshift.

\section {SFRs from adding UV and IR emission}
\label{Lir.sec}

%%%%%%%
% FIG 1
%%%%%%%
\begin {figure}[htbp]
%\vspace{0.2in}
\centering
\epsscale{1.0}
\plotone{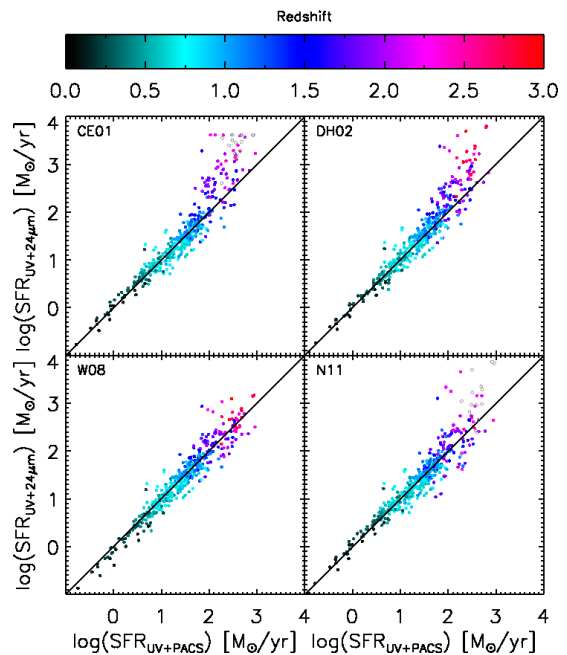} %{figeps/comp_SFR_24IR4.eps}
%\epsscale{1.0}
\caption{
Star formation rates of galaxies at $0 < z < 3$ based on UV plus 24
$\mu$m emission versus UV plus PACS emission.  The IR part of the SFR
is proportional to the total IR luminosity $L_{IR}$, which was derived
monochromatically using conversions and template sets from CE01, DH02,
W08, and N11.  Adopting the locally calibrated luminosity-dependent
recipes from CE01 and DH02 to convert observed 24 $\mu$m fluxes to
$L_{IR}$ leads to overpredicted SFRs at the high-redshift and high-SFR
end.  For N11, these cases are limited to redshifts above $z > 2.5$, where the 24 $\mu$m filter enters $\lambda_{rest} < 6\ \mu$m, and the CE01/N11 templates lack a physically realistic calibration (open gray circles).
\label{24IR.fig}}
\end {figure} 

%%%%%%%
% FIG 2
%%%%%%%
\begin {figure}[htbp]
\centering
%\epsscale{0.55}
\plotone{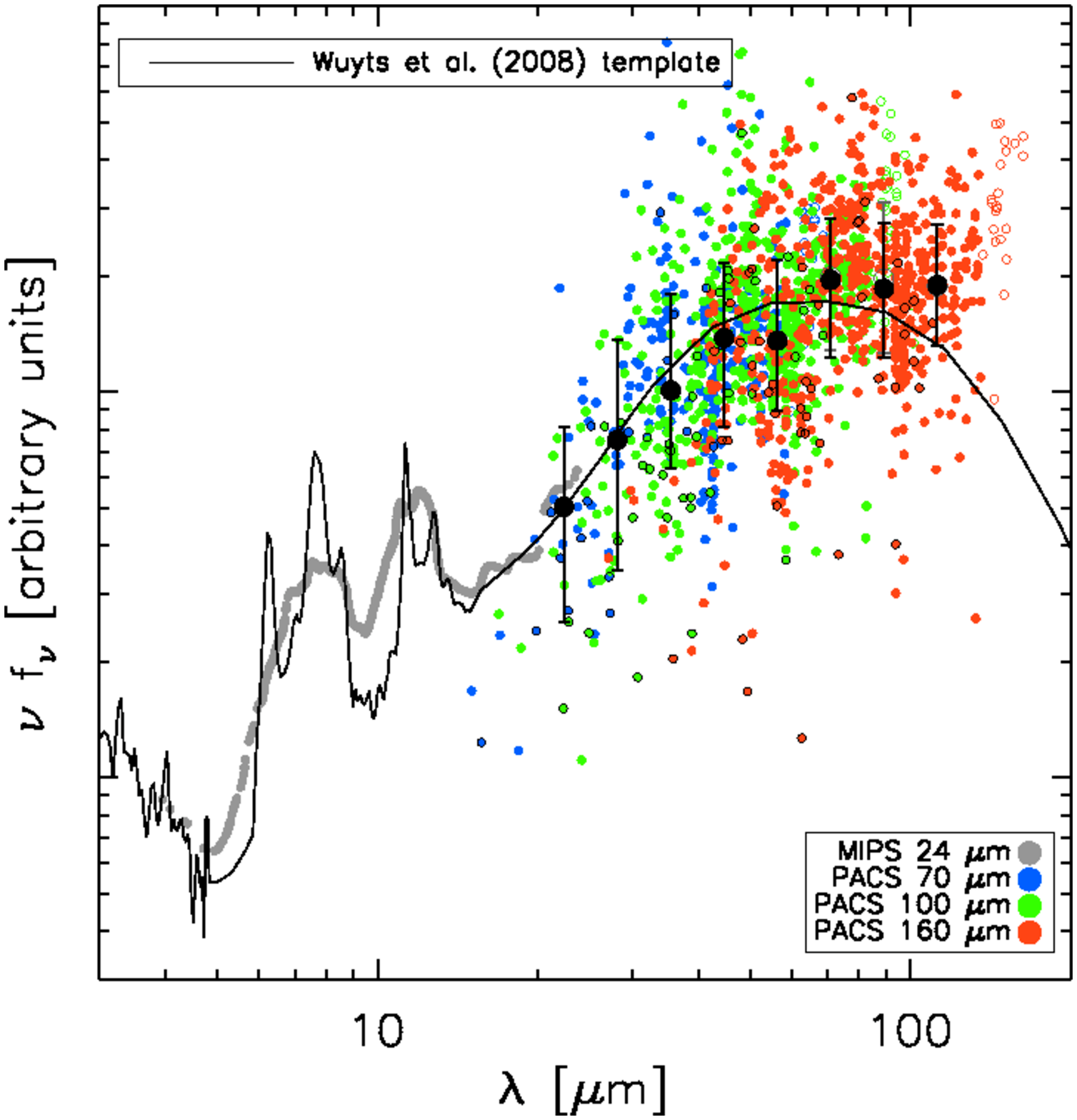} %{figeps/template.eps}
\vspace{0.1in}
\plotone{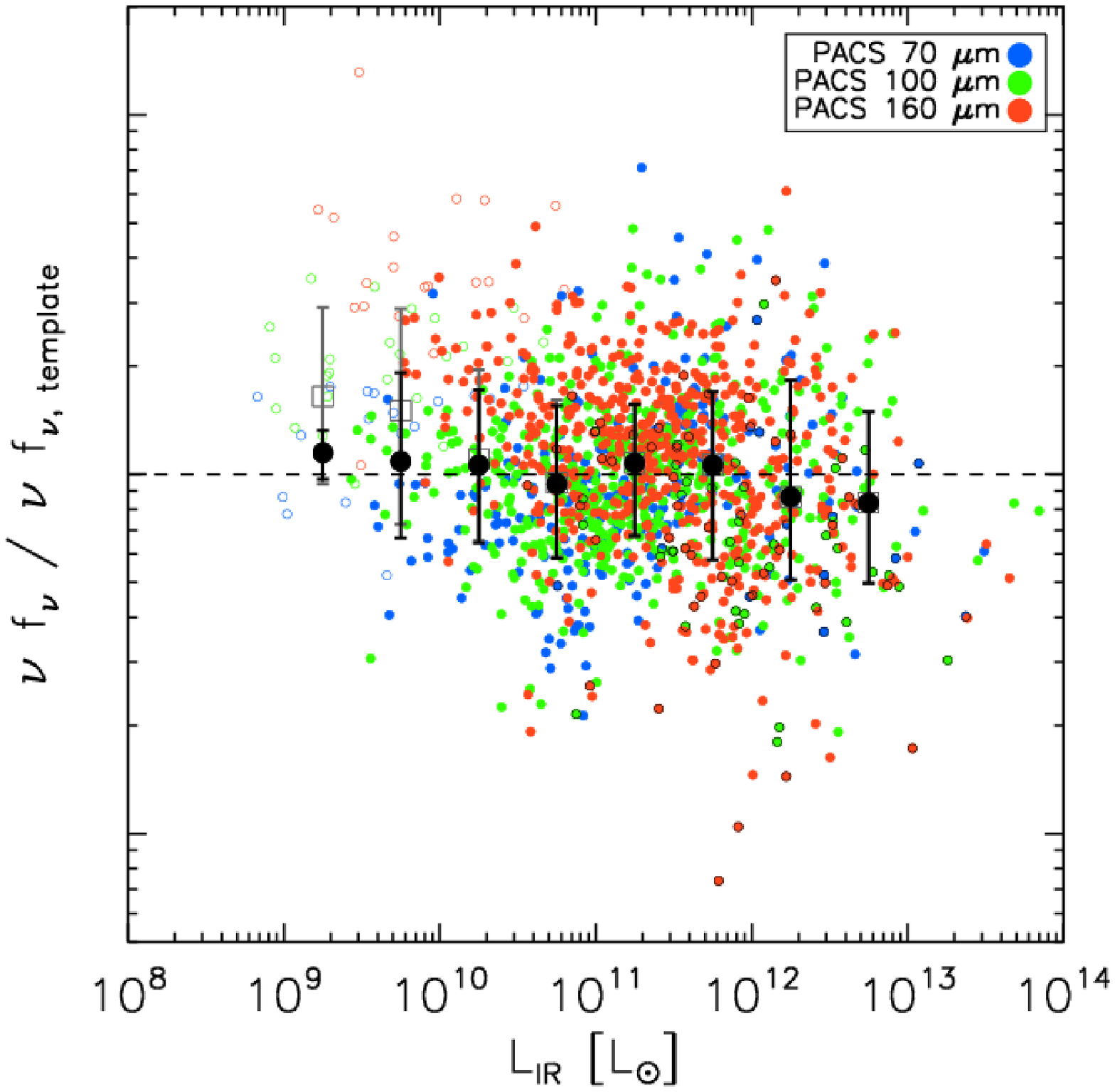} %{figeps/Ldependence.eps}
%\epsscale{1.0}
\caption{{\it Top:} Spectral energy distribution template by Wuyts et al. (2008), with the observed photometry of PACS-detected galaxies at $z < 3$ in the GOODS-South field overplotted at the respective rest-frame wavelengths.  For every galaxy, a uniform scaling is applied to its $24 - 160\ \mu$m photometry so as to normalize 
the observed 24 $\mu$m flux ({\it gray}) to the template ({\it black}) convolved by the MIPS passband.  Large black symbols indicate the binned median and standard 
deviation of the scaled PACS photometry, matching the template, albeit with a significant (0.25 dex) scatter.  Small black circles mark sources with an X-ray luminosity 
above $10^{42}\ erg\ s^{-1}$.
{\it Bottom:} Residual from the Wuyts et al. (2008) template as function of inferred total IR luminosity.  Large empty boxes and filled circles indicate the binned median, with and without 
galaxies at $z < 0.2$ respectively.  Excluding the most nearby galaxies (small
empty circles), only a marginal trend with IR luminosity remains.
\label{Lir.fig}}
\end {figure}

In this Section, we analyze SFR indicators based on adding
the unobscured and obscured star formation, traced by the UV and IR
emission respectively.  With the advent of PACS/Herschel, the emission
at or near the IR peak can now be probed directly out to high redshift
to unprecedented depths (see, e.g., Nordon et al. 2010; Elbaz et
al. 2010).  Combining PACS and SPIRE data, Elbaz et al. (2010) found
that the total IR luminosity ($L_{IR} \equiv L(8 - 1000\ \mu m)$)
can be derived reliably from a single PACS band.  We will therefore
restrict ourselves to such monochromatic conversions to $L_{IR}$ in
this paper.

The depth of PACS imaging in the GOODS-South field (2.0 mJy at 160
$\mu$m, 3$\sigma$) is among the deepest of all IR lookback surveys.
In most other fields, the derivation of $L_{IR}$ or $SFR_{UV+IR}$ from
PACS photometry has to rely on stacking analyses to reach the same
depths.  In order to preserve information on an individual object
level, IR luminosities for all but the brightest galaxies must then be
derived from MIPS 24 $\mu$m.  Here, we exploit the deep mid- and
far-IR data in GOODS-South to test for internal consistency between them.

We consider four conversion recipes.  The conversions by Chary \&
Elbaz (2001, hereafter CE01) and Dale \& Helou (2002, hereafter DH02)
rely on locally calibrated template libraries, where the template
adopted from the library depends on the galaxy's luminosity (generally
yielding larger conversion factors for more luminous galaxies at a
given redshift).  Studying PACS observations in GOODS-North, Nordon et
al. (2010) argued for enhanced emission of polycyclic aromatic
hydrocarbons (PAHs) in $z \sim 2$ galaxies relative to local galaxy
SEDs of the same $L_{IR}$.  Accounting for this, Nordon et al. (2011,
hereafter N11) formulate a recalibration of the CE01 library for the
$1.5 < z < 2.5$ regime (below $z < 1.5$, the original CE01 conversion
is used).  Specifically, these authors find little to no luminosity
dependence of the rest-frame 20-60 $\mu$m SED shape for the $L_{IR}
\gtrsim 10^{11}\ L_{\sun}$ regime probed.  In the mid-infrared, their
conversion has a luminosity dependence, but such that the ratio $\nu
L_{\nu}(rest-8\mu m) / L_{IR}$ at a given redshift and luminosity is
larger than that from CE01.  Finally, Wuyts et al. (2008) introduced a
luminosity-independent conversion based on a single template, which
was motivated by MIPS 24 $\mu$m, 70 $\mu$m, and 160 $\mu$m stacking
results by Papovich et al. (2007).  This template was constructed by
averaging the logarithm of DH02 templates with the parametrization
$\alpha$ of the intensity of the interstellar radiation field in the
range $1 < \alpha < 2.5$ in steps of 0.0625. \footnote{The template,
  as well as a table with conversion factors from MIPS 24 $\mu$m, and
  PACS 70 $\mu$m, 100 $\mu$m and 160 $\mu$m to $L_{IR}$ is released at
  http://www.mpe.mpg.de/\textasciitilde swuyts/Lir\_template.html} In
terms of local analogs, its mid- to far-infrared SED shape is more
reminiscent of M82 than of Arp220.

In Figure\ \ref{24IR.fig}, we present a comparison of $SFR_{UV+24\mu
  m}$ to $SFR_{UV+PACS}$ for PACS-detected galaxies out to $z = 3$,
for each of the above $L_{IR}$ conversion recipes.  In each case, the
total SFR is derived following Kennicutt (1998):

\begin{equation}
SFR_{UV+IR}\ [M_{\sun}\ yr^{-1}] = 1.09 \times 10^{-10}\ (L_{IR} + 3.3
L_{2800}) / L_{\sun}
\label{SFR_UVIR.eq}
\end{equation}

where the relative scaling of the UV and IR contribution is based on
calibrations in the local universe, and the overall scaling factor
assumes a Chabrier (2003) IMF.  The rest-frame luminosity $L_{2800}
\equiv \nu L_{\nu}(2800\AA)$ was computed with EAZY from the
best-fitting SED, itself a superposition of the 6 EAZY
principle-component templates.  Only at the lowest redshifts ($z <
0.3$), rest-frame 2800\AA\ falls blueward of the available photometry.
For $SFR_{UV+PACS}$, we use the longest wavelength PACS band that had
a significant ($> 3\sigma$) detection.  The color-coding indicates the
redshift of the galaxies.  It is a well-known fact that galaxies that
are highly actively star-forming are more common at larger lookback
times.  The lower end of the SFR distribution at a given redshift is
simply determined by the flux limit of our sample.

At $z \ll 2$ and $SFR < 100\ M_{\sun} / yr$, we find little difference
between the PACS and 24 $\mu$m-based estimates, irrespective of the
conversion recipe used.  However, above $z \gtrsim 2$ and for SFRs above
$100\ M_{\sun} / yr$, the 24 $\mu$m-based estimates using the locally
calibrated CE01 and DH02 conversions are clearly overestimated with
respect to those based on PACS photometry.  At $1.5 < z < 2.5$, no
such systematic bias is present in the case of the W08 and N11
conversions.  These results imply that, rather than having mid- to
far-IR SEDs as local ULIRGs, ULIRGs at $z \sim 2$ are better
characterized as scaled-up versions of local LIRGs, which are colder
than nearby ULIRGs.  This finding is in agreement with recent results
by Nordon et al. (2010), Elbaz et al. (2010), Muzzin et al. (2010),
and Symeonidis et al. (2011).

We point out that any variation in $\nu L_{\nu}(24\mu m) / L_{IR}$,
where $L_{\nu}(24\mu m)$ refers to the specific luminosity probed by
the observed 24 $\mu$m band, translates superlinearly to variations in
$L_{IR}(24\ \mu m) / L_{IR}(160\ \mu m)$ when applying a
luminosity-dependent conversion.  This leads to a somewhat increased
scatter around unity in the $L_{IR}(24\ \mu m) / L_{IR}(160\ \mu m)$ ratio of $1.5
< z < 2.5$ galaxies for N11 ($\sim 0.4$ dex) compared to W08 ($\sim
0.3$ dex), since the scatter in $\nu L_{\nu}(24\mu m) / L_{IR}$ at a
given luminosity is large compared to any systematic variation of $\nu
L_{\nu}(24\mu m) / L_{IR}$ over the luminosity range probed.  More
importantly, the luminosity-independent conversion by W08 has the
advantage of being applicable also above $z > 2.5$, where N11 shows
more systematic outliers.  This is at least in part due to the fact
that the 24 $\mu$m band starts entering the $\lambda_{rest} \lesssim
6\ \mu$m regime where the CE01 templates lack a physically realistic
calibration.

In order to address the uncertainties involved in converting 24 $\mu$m
to $L_{IR}$ in more detail, we present the W08 template in Figure\
\ref{Lir.fig}.  Overplotted, we show the IR photometry of $0 < z < 3$
galaxies that are individually detected ($> 3\sigma$) in at least one
PACS band.  Their 24 $\mu$m flux is normalized to match the template
flux within the passband at the respective redshift, but their 24
$\mu$m - PACS colors are left as observed.  Black filled circles and
error bars mark the median and central 68th percentiles for
PACS-detected galaxies at $0.2<z<3$, respectively.  Gray empty circles
and error bars illustrate small changes when including the lowest
redshift ($z<0.2$) sources.  Although a significant scatter (0.25 dex
in $\Delta \log \nu f_{\nu}$) is notable, the observations are in the
median consistent with the W08 template.  Moreover, Figure\
\ref{Lir.fig}b demonstrates that the residuals from the template do
not strongly correlate with the $L_{IR}$ derived from the respective
PACS band, especially when excluding local ($z < 0.2$) galaxies, that
dominate the lowest luminosities ($10^9\ L_{\sun} < L_{IR} < 10^{10}\
L_{\sun}$).  In other words, we observe a spread in 24 $\mu$m - PACS
colors, even at a given redshift, but this spread does not seem to
correlate strongly with galaxy luminosity.  As a consequence of the
observed flux limits, different luminosity bins in Figure\
\ref{Lir.fig}b will naturally be populated by galaxies of a different
redshift distribution.  This could potentially mask some luminosity
dependence present within a narrow redshift slice.  Including also
observations at 16 $\mu$m, N11 investigate this by probing the
rest-frame 8 $\mu$m luminosity in bins of both $L_{IR}$ and redshift,
finding a dependence on both albeit with significant scatter.  While
not capturing these physics, the W08 SED can be considered as an
effective template offering a simple and robust conversion from
observed 24 $\mu$m photometry to $L_{IR}$.  The 7.7 $\mu$m PAH
strength may increase with redshift relative to $L_{IR}$ (see N11),
but for the purpose of converting MIPS 24 $\mu$m to $L_{IR}$, it is
sufficient that its strength in the template is appropriate for $z
\sim 2$, where this feature is sampled by the 24 $\mu$m passband.
Given that the scatter in $\nu L_{\nu}(24\mu m)/L_{IR}$ is large
compared to any variation with luminosity, not accounting for a
luminosity dependence has the compensating advantage that deviations
in $\nu L_{\nu}(24\mu m)$ at a given luminosity do not translate
nonlinearly to $L_{IR}(24\mu m)$.  Another parameter potentially
contributing to the scatter observed in Figure\ \ref{Lir.fig}b, is the
dust temperature.  More specifically, variations in the dust
temperature from galaxy to galaxy would express themselves as
deviations from a unique SED shape, and hence a source of scatter in
Figure\ \ref{Lir.fig}b.  Magdis et al. (2010) recently reported a wide
range of dust temperatures in high-redshift ULIRGs.  For an in depth
discussion of the relation between dust temperatures and other
characteristics of high-redshift galaxies (such as SFR and mass), we
defer the reader to Magnelli et al. (in prep).  In the Appendix, we
briefly discuss possible contributions by Active Galactic Nuclei
(AGN), but argue that they do not affect the above results for the
ensemble of PACS-detected galaxies.  For a first exploration of AGN in
the context of PACS observations, we refer the reader to Shao et
al. (2010).

In the above analysis, we focussed on PACS detections only.  While
half of the 24 $\mu$m sources in GOODS-South are not detected at any
PACS wavelength, the PACS sample is matched quite well to the 24
$\mu$m depth of other, shallower fields.  For example, 90\% of the
galaxies brighter than the 24 $\mu$m detection threshold of COSMOS are
PACS-detected in GOODS-South.  This implies that the conversion
described in this section can safely be applied to translate the
COSMOS 24 $\mu$m photometry to $L_{IR}$.  We briefly explored how the
median SED shape changes when going beyond the PACS detection
threshold by means of stacking.  We computed the median stacked fluxes
in bins of rest-frame wavelength now including all sources detected at
24 $\mu$m.  Stacking was performed after scaling the PACS postage
stamps by the same factor used to normalize the 24 $\mu$m flux of the
corresponding object to the W08 template.  Postage stamps of
PACS-undetected sources were cut out of residual maps from which all
individually detected sources were subtracted.  We find that the blue
side of the far-IR bump ($20\ \mu m < \lambda_{rest} < 60\ \mu m$) is
somewhat more depressed, by $\sim 0.15$ dex with respect to the W08
template to which the 24 $\mu$m flux was normalized.  However, the
amplitude of the peak in IR emission, where the bulk of the energy
emitted at 8 - 1000 $\mu$m comes from, is well matched by the
template.  The 8 - 1000 $\mu$m integral over the DH02 or CE01 template
that best fits the shape of the median stacked SED deviates by merely
a percent from that of the W08 template, making the latter an
acceptable basis for deriving $L_{IR}$ for our full 24 $\mu$m-detected
sample.

We conclude that a continuity between 24 $\mu$m and far-infrared based
SFR indicators can be obtained out to $z = 3$, albeit with significant
scatter.  The good median performance of the W08
conversion in the light of new Herschel observations is encouraging
for recent studies in which it was applied (Franx et al. 2008; Wuyts
et al. 2009b; Damen et al. 2009; Kajisawa et al. 2010).

\section {SFRs from SED Modeling and dust-corrected UV emission}
\label{SFR_SED.sec}

In GOODS-South, not only the PACS but also the MIPS depth (0.02 mJy at 24 $\mu$m, 5$\sigma$) is
one of the deepest available.  Consequently, in studies that exploit larger
areas to reduce cosmic variance effects, one will be limited by the
shallower MIPS imaging of other fields.  Other SFR tracers than
IR-based methods are therefore indispensable.

\subsection{Cross-calibration to $SFR_{UV+IR}$}
\label{SEDvsUVIR.sec}

Modeling of broad-band SEDs with stellar population synthesis models
is one of the most commonly used approaches, as it is a more generally accessible
means to derive SFRs (among other stellar population properties) for
galaxies over a wide range of redshifts and types.  It can also be
extended to fainter galaxies with lower SFRs.

We investigate the performance of this technique, which is essentially
a dust correction method, by comparing it to $SFR_{UV+IR}$ for
galaxies with a PACS and/or MIPS detection.  The $L_{IR}$ that enters
the computation of $SFR_{UV+IR}$ was calculated using the W08 template
and the longest wavelength IR band that had a significant detection.
We adopt default assumptions regarding SED modeling.  Briefly,
exponentially declining SFHs, also known as $\tau$ models, were
constructed from the Bruzual \& Charlot (2003, hereafter BC03) stellar
population synthesis models, and fitted to the $U$-to-8 $\mu$m
FIREWORKS SEDs using the fitting code FAST (Kriek et al. 2009).  We
are aware that increasing star formation histories have recently been
proposed as more appropriate for galaxy evolution at early times ($z
\gtrsim 3$, Renzini et al. 2009; Maraston et al. 2010; Papovich et
al. 2011), and explore this option further in Section\
\ref{delayedtau.sec}.  For now, we restrict the analysis of our $0 < z
< 3$ sample to default $\tau$ models whose wide use in the literature
makes it worthwhile testing their performance in light of the new IR
constraints.  As throughout the paper, we adopt a Chabrier (2003) IMF,
and we assume solar metallicity.  We required the time since the onset
of star formation to be at least 50 Myr, and not exceeding the age of
the universe at the epoch of observation.  We experimented with
minimum e-folding times varying between 10 Myr and 10 Gyr.  The fitting
procedure minimizes the $\chi ^2$ statistic taking discrete steps of
0.1 dex in age and $\tau$ within the allowed range.  We allow the
galaxies to be attenuated within a range of visual extinctions $0 <
A_V < 4$, in steps of 0.1 mag, assuming a uniform foreground screen
geometry, and with the reddening following the Calzetti et al. (2000)
law.  We refer to the instantaneous SFR of the best-fit model as
$SFR_{SED}$.

%%%%%%%
% FIG 3
%%%%%%%
\begin {figure}[t]
\centering
%\epsscale{0.8}
\plotone{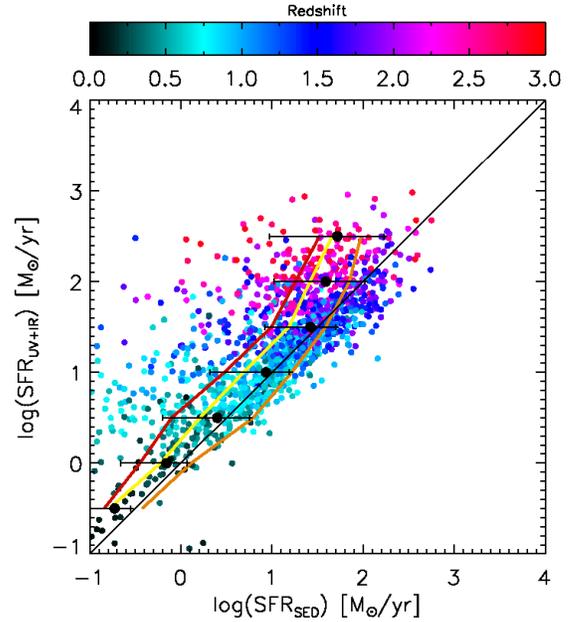} %{figeps/comp_SFRbasic.eps}
%\epsscale{1.0}
\caption{The SFRs of IR-detected galaxies at $0 < z < 3$ based on
  adding the unobscured (UV) and re-emitted (IR) radiation of newly
  formed stars versus SFRs from stellar population modeling of observed
  optical-to-mid-IR SEDs.  Here, SFR$_{\rm SED}$ was derived using
  BC03 models with exponentially declining SFHs (with $\log
  (\tau_{min}) = 8.5$).  For modest levels of star formation ($SFR
  \lesssim 50\ M_{\sun}\ yr^{-1}$), both SFR indicators agree well in
  the median (large black circles, with error bars indicating the
  central 68th percentile).  At higher SFRs, which become increasingly more common
  toward higher redshift, the SED-modeled SFR systematically
  underpredicts that inferred from UV + IR.  Red and yellow curves
  illustrate the median relation when adopting less stringent
  constraints ($\log (\tau_{min}) = 7$ and 7.5 respectively) on the
  minimum e-folding time.  The orange line marks the median relation
  when allowing very long ($\log(\tau_{min})=10$) e-folding times only.
\label{SFRbasic.fig}}
\end {figure} 

We find the best correspondence to $SFR_{UV+IR}$ when forcing the
e-folding time to be larger than $\log \tau_{min} = 8.5$.  The plotted
points in Figure\ \ref{SFRbasic.fig} correspond to this set of
boundary conditions.  The majority of galaxies, at low to intermediate
($SFR_{UV+IR} \lesssim 50\ M_{\sun}/yr$) SFRs line up along a ridge
line which coincides with the one-to-one relation.  A tail toward
lower SED-modeled SFRs is present, and contains more than 50\% of the
galaxies at the high SFR (and high redshift) end.  When allowing
shorter e-folding times, we find these are often preferred in a
least-squares sense, but leave a systematic offset in SFR in
comparison to measurements of $SFR_{UV+IR}$.  The red and yellow line
in Figure\ \ref{SFRbasic.fig} indicate the median trend for $\log \tau
_{min} = 7.0$ and 7.5 respectively.  It thus seems that e-folding
timescales of at least several 100 Myr give results that are more consistent
with UV+IR based estimates than more rapidly declining
histories.\footnote{If we were to average the best-fit SFH over the
  last 100 Myr instead of adopting the instantaneous SFR, which for
  $\tau$ models implies an increase by a factor $\frac{\tau}{100Myr}
  \left ( e^{\frac{\tau}{100Myr}} - 1 \right )$, the dependence on
  $\tau_{min}$ reduces, and SFRs derived with short $\tau_{min}$ would
  agree better with $SFR_{UV+IR}$.  However, given the implausibly
  young ages associated with those fits (see Section\
  \ref{tracingback.sec}) longer e-folding times still seem preferred.}
In Section\ \ref{tracingback.sec}, we will describe independent
evidence in this direction, based on the implausibly young ages
inferred from fits with small $\tau$ values.  Changing
$\log(\tau_{min})$ from 8.5 to 9 leads to a similar performance in the
comparison to $SFR_{UV+IR}$.  However, when allowing very long
e-folding times only ($\log(\tau_{min})=10$), $SFR_{SED}$ becomes an
overestimate at intermediate star formation levels (see orange curve
in Figure\ \ref{SFRbasic.fig}).  Overall, we recommend
$\log(\tau_{min})=8.5$, for its effectiveness at reproducing the low to
intermediate SFRs while not overly restricting the allowed range of SFHs.

To contrast this timescale of $\sim 300$ Myr for variations in star
formation activity with another characteristic timescale of the
galaxies in our sample, we computed the dynamical time $t_{dyn} =
\sqrt{ \frac {R^3}{2GM} }$ of the 27\% of galaxies that are covered by
the HST/WFC3 Early Release Science program.  Here, we measured the
effective radius $R$ by fitting Sersic models to the $H$-band surface
brightness distribution using GALFIT v3.0 (Peng et al. 2010).  We find
typical $\tau / t_{dyn}$ ratios of order $\sim 50$, with large
variations from 10 to 100 and up.  This implies that most star
formation happened in a relatively stable mode over long timescales,
as opposed to a single short burst taking place on a dynamical time.
Similar conclusions were drawn by Genzel et al. (2010), who pushed the
study of the Schmidt-Kennicutt law (Schmidt 1959; Kennicutt 1998) and
the Elmegreen-Silk relationship (Elmegreen 1997; Silk 1997) for normal
star-forming galaxies to $z \sim 1 - 3$.

%%%%%%%
% FIG 4
%%%%%%%
\begin {figure}[t]
\centering
%\epsscale{0.8}
\plotone{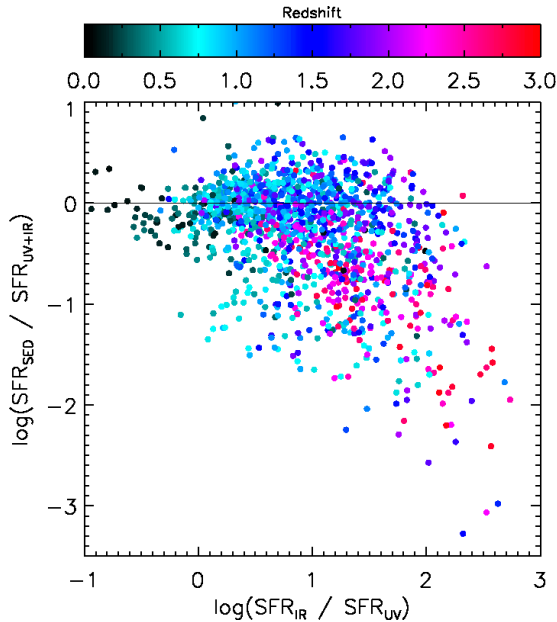} %{figeps/Del_vs_IRoverUV.eps}
%\epsscale{1.0}
\caption{Ratio between SFR$_{\rm SED}$ and SFR$_{UV+IR}$ as function of relative fraction of radiation from young stars that is re-processed by dust versus 
escaping without attenuation.  The underprediction of the total amount
of star formation by SFR$_{\rm SED}$ (i.e., a dust correction method)
is most severe for galaxies that have the largest $SFR_{IR}/SFR_{UV}$
ratio.  These conditions are more prevalent in vigorously star-forming
systems, as found more commonly at high redshift.  Saturation of
reddening as an extinction tracer due to a patchy dust distribution is responsible for this effect.
\label{Del_vs_IRoverUV.fig}}
%\vspace{-0.2in}
\end {figure} 

Despite the good overall correspondence between the SFR indicators,
the tail of systematic underestimates by $SFR_{SED}$, reaching an
order of magnitude at the largest $SFR_{UV+IR} \sim 300\ M_{\sun}/yr$
(predominantly galaxies at $z \gtrsim 2.5$) remains worrisome.  Such a
trend at the high-SFR end was previously signaled by Santini et
al. (2009) on the basis of pre-Herschel data sets.  Figure\
\ref{Del_vs_IRoverUV.fig} sheds light on its origin.  The deviation
from a one-to-one relation between $SFR_{UV+IR}$ and $SFR_{SED}$ is a
clear function of the relative ratio of emission from young stars that
is re-processed by dust versus escaping unhindered as UV light.  The
diagram illustrates that particularly when the unattenuated UV
emission reveals only a few percent of the total star formation (i.e.,
when the dust correction factor is very large), a simplistic dust
correction method based on a uniform foreground screen does not
account properly for optical depth effects.  The reason is that
information on the dust correction factor is not directly embedded in
the galaxy's SED.  It is the reddening of the spectral slope that is
directly traced, but only translates linearly to extinction when the
obscuring material has the configuration of a foreground screen.  In
case of patchy dust obscuration, or dust that is mixed with the
emitting sources, reddening saturates as a tracer of extinction.  In
the case of selective extinction towards young star-forming regions,
the resulting underestimate of the SFR can be even more pronounced
(Poggianti et al. 2001; Wuyts et al. 2009a).  The same effect is seen
in local ULIRGs (Goldader et al. 2002) and starburst galaxies (e.g.,
F\"{o}rster Schreiber 2001 and references therein) that lie above the
well-known IRX-$\beta$ relation (Meurer et al. 1999), and in
hydrodynamic simulations of high-redshift (merging) galaxies (Wuyts et
al. 2009a, Fig. 10).  Our conclusion confirms the recent results by
Reddy et al. (2010), who find that galaxies with large IR luminosities
($L > 10^{12}\ \L_{\sun}$) at $z \sim 2$ also lie above the
IRX-$\beta$ relation, unlike more modestly star-forming galaxies at
the same epoch.

\subsection{Single-color SFRs for Star-forming $z \sim 2$ Galaxies}
\label{daddi.sec}

%%%%%%%
% FIG 5
%%%%%%%
\begin {figure}[t]
\centering
%\epsscale{0.8}
\plotone{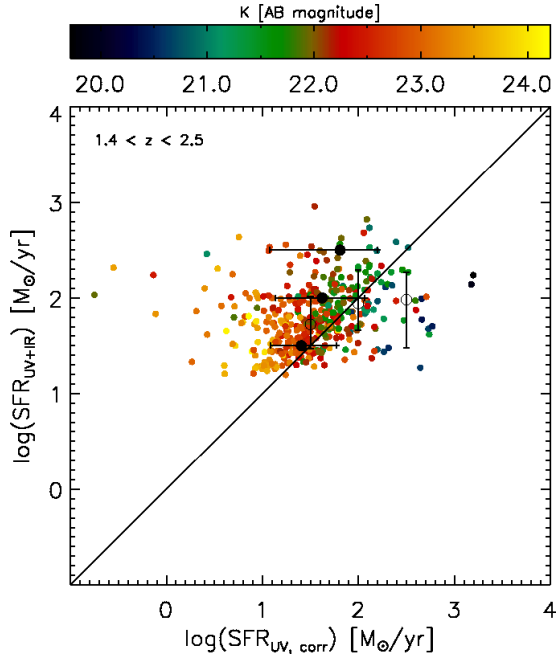} %{figeps/comp_SFR_daddi.eps}
%\epsscale{1.0}
\caption{
The SFRs of IR-detected galaxies at $1.4 < z < 2.5$ based on adding
the unobscured (UV) and re-emitted (IR) radiation of newly formed
stars versus the Daddi et al. (2007) dust correction method which
relies solely on the observed $B - z$ color.  Large filled and open
symbols indicate the median in bins of $SFR_{UV+IR}$ and
$SFR_{UV,corr}$ respectively.  Whereas the two methods yield
consistent results at modest SFR ($ \lesssim 50\ M_{\sun}\ yr^{-1}$)
levels, the dust correction method fails to recover the total amount
of star formation at the highest SFR$_{UV+IR}$ bin.  Note that
inferred systematic offsets are dependent on the K-band limit of the
sample.
  \label{SFRdaddi.fig}}
%\vspace{-0.2in}
\end {figure} 

Daddi et al. (2004; 2007a) introduced an observed color criterion to
dust-correct UV SFRs of BzK galaxies at $1.4 < z < 2.5$.  Since it
is based on observed-frame, rather than rest-frame photometry, it has the
limitation of only being applicable to the redshift range for which it
was designed.  On the other hand, it has the advantage of being
more easily accessible, as it only requires $B$ and $z$ photometry (and
$K$ in order to select the high-redshift galaxies in the first place).
Furthermore, it has been argued that the use of such a limited
wavelength range avoids dilution of the information on the rate of
star formation by degeneracies with other stellar population
parameters.  It is therefore useful to address the performance of the
Daddi et al. (2007a) criterion in light of the $SFR_{UV+IR}$
calibrations (Section\ \ref{Lir.sec}), that are now motivated by the
new Herschel/PEP photometry.

Figure\ \ref{SFRdaddi.fig} presents a comparison of $SFR_{UV+IR}$ to
the single-color dust correction method by Daddi et al. (2007).  We
plot all galaxies with an IR detection (i.e., for which $SFR_{UV+IR}$
is not an upper limit), most of which satisfy the BzK criterion.
Despite the limited range in redshift and SFR probed, a similar trend
emerges as seen for the SED modeling (Section\ \ref{SEDvsUVIR.sec}).
While at intermediate $SFR_{UV+IR}$, the two SFR indicators agree
well, the dust correction method fails to recover the total amount of
star formation at the largest $SFR_{UV+IR}$.  We note that, when
binning in $SFR_{UV, corr}$ (i.e., the SFR indicator to be tested)
instead of $SFR_{UV+IR}$ (our reference SFR indicator), the median
$SFR_{UV+IR}$ of the highest $SFR_{UV, corr}$ bin actually lies below
the one-to-one relation ({\it large open circles}).  The fact that the
median binned relation depends on the axis that is binned reflects a
significant scatter in the relation between the UV+IR and
dust-corrected SFR.  The absence of a tight relation may stem from
variations in the dust distribution from galaxy to galaxy.  The
color-coding in Figure\ \ref{SFRdaddi.fig} furthermore indicates that
the scatter is not random, but that instead the relation between the
two SFR indicators is a function of $K$-band magnitude.  This
illustrates that the offset with respect to $SFR_{UV+IR}$ depends on
the properties of the galaxy sample.  E.g., stacking $K$-bright ($K <
22$) galaxies in bins of $SFR_{UV, corr}$ Nordon et al. (2010) found
an excess of 0.3 dex for dust-corrected UV SFRs.

We conclude that we do not find any evidence that a single color dust
correction method would be more precise than SFRs from modeling of the
full $U$-to-8 $\mu$m SED.  Both dust correction methods perform
equally well in the redshift range $1.4 < z < 2.5$ where they overlap,
with similar biases for the dusty, most rapidly star-forming systems.
Resolved SED modeling on the scale of the patchiness of the dust
distribution may help to overcome the saturation of reddening as an
extinction tracer when IR data are not available (see Section\
\ref{ages.sec}, and in more detail Wuyts et al. in prep).

%test this explicitly SFR_UV+IR vs SFR_SED for 1.4 < z < 2.5

\subsection{Dependence on Stellar Population Synthesis Models}
\label{stelpop.sec}

%%%%%%%
% FIG 6
%%%%%%%
\begin {figure}[t]
\centering
%\epsscale{0.9}
\plotone{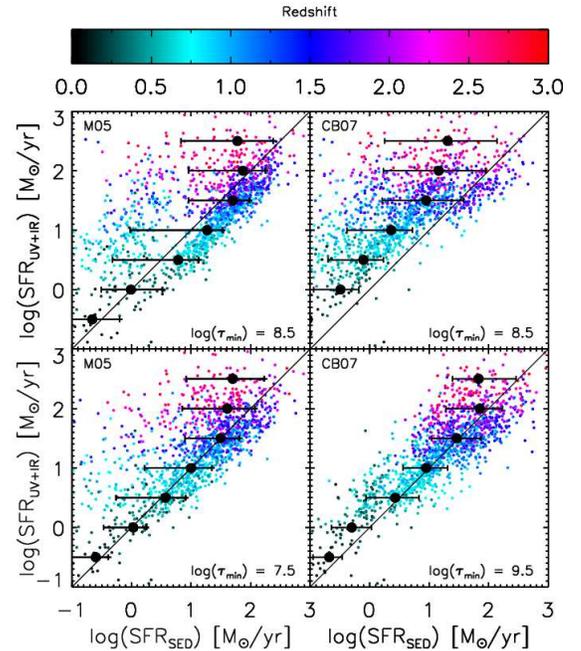} %{figeps/comp_SFRstelpop.eps}
%\epsscale{1.0}
\caption{Idem as Figure\ \ref{SFRbasic.fig}, but using different stellar population synthesis codes (M05 and CB07 instead of BC03) to derive SFR$_{\rm SED}$.  In the 
top panels, identical constraints on the SFH (specifically the minimum allowed e-folding time $\tau_{min}$) as for Figure\ \ref{SFRbasic.fig} were used.  This leads to 
SED-modeled SFRs that are more discrepant from SFR$_{UV+IR}$ than SFR$_{\rm SED, BC03}$.  By tuning $\tau_{min}$ (bottom panels), a 
correspondence to SFR$_{UV+IR}$ of similar quality can be obtained.  In all cases, the saturation of reddening as extinction tracer in dusty, star-bursting (high-redshift) galaxies leads to systematic underestimates of SFR by the SED modeling method.
\label{SFRstelpop.fig}}
\end {figure} 

Even under identical assumptions on SFH and dust reddening,
differences between stellar population synthesis codes give rise to
systematic uncertainties in the determination of stellar population
properties.  The impact of the treatment of certain stellar
evolutionary phases, such as the thermally pulsating AGB phase, on the
estimated stellar mass and age of (high-z) galaxies has been discussed
extensively elsewhere (see, e.g., Maraston 2005, hereafter M05;
Maraston et al. 2006; Wuyts et al. 2007).  In Figure\
\ref{SFRstelpop.fig}, we focus on the implications for SFR recipes.
The top two panels compare $SFR_{UV+IR}$ to $SFR_{SED}$ as derived
with M05 and Charlot \& Bruzual (2007, hereafter CB07) models.  We
adopted the same restriction of a minimum e-folding time of $\log
\tau_{min} = 8.5$ that provided the best agreement for BC03 models.
Similar to the BC03 comparison in Figure\ \ref{SFRbasic.fig}, the SED
modeled SFRs lead to underestimates at the high $SFR_{UV+IR}$ end.
However, at intermediate and low SFRs the behavior looks markedly
different, especially for the CB07 models that lead to consistently
lower estimates of the SFR by $\sim 0.4$ dex.  Adjusting a single
parameter $\tau_{min}$ in our SED modeling recipe, it is possible to
calibrate both $SFR_{SED,\ M05}$ and $SFR_{SED,\ CB07}$ such that they
produce an equally good match to $SFR_{UV+IR}$ as $SFR_{SED,\ BC03}$.
In other words, our cross-calibration of SFR indicators cannot
discriminate between different stellar population synthesis models.
We find, however, that BC03, M05 and CB07 models require different
constraints on the minimum e-folding time ($\log \tau_{min} = 8.5$,
7.5, and 9.5 respectively) in order to optimally match the UV+IR based
indicator.  

Figure\ \ref{SFRstelpop.fig} demonstrates that the SED fitting results
based on M05 and CB07 models show qualitative differences despite the
fact that both contain an updated implementation of the TP-AGB phase,
leading to increased rest-frame $K$-band fluxes and lower stellar mass
estimates.  The two stellar population synthesis models differ in at
least two aspects.  First, the boost in rest-frame $K$-band light due
to TP-AGB stars occurs at earlier times for CB07 (around $\log age
\sim 8.5$) compared to M05 (around $\log age \sim 9$).  Second, the
optical (rest-frame $U-V$) colors by CB07 are identical to those of
BC03, whereas they are redder by nearly 0.2 mag in the M05 models for
most of the ages considered.  This stems from the use of different
stellar tracks that do (in the case of CB07) or do not (for M05)
include convective overshooting (see Maraston et al. 2006; Fagotto et
al. 1994 for the Padova stellar tracks used by CB07; Cassisi et
al. 1997 for the Frascati stellar tracks used by M05).  Together,
these differences lead to best-fit solutions that in the case of M05
correspond to younger, dustier and more actively star-forming systems
than inferred from BC03 fits, while CB07 fits yield lower SFRs and
$A_V$, and older ages.  As we will discuss in Section\
\ref{consistency.sec}, the stellar ages derived from BC03 models
without stringent constraints on formation redshift or e-folding time
tend to be implausibly young.  This problem may therefore be even more
pressing when considering M05 fits.  Finally, we note that for both
M05 and CB07 fits, solutions with ages for which the TP-AGB
contribution is maximal, seem to be avoided.  The apparent bimodality
in the upper left panel of Figure\ \ref{SFRstelpop.fig} is a
reflection of this bimodality in estimated ages.  Along similar lines,
Kriek et al. (2010) recently reported that, for their sample of
post-starburst galaxies at $0.7 < z < 2.0$, a low contribution from
TP-AGB stars was demanded in order to reproduce their spectral energy
distributions.

To conclude, we find that, in common for all stellar population models
considered, underestimates of the SFR for the dusty, most actively
star-forming galaxies, that are abundantly present at high redshift,
are intrinsic to any dust correction method assuming a uniform
foreground screen.  SED modeling recipes can be tuned to reproduce the
level of star formation in galaxies with more modest $SFR_{UV+IR}$.

\subsection{Dependence on Extinction Curve}
\label{extinction.sec}

So far, we always assumed the Calzetti et al. (2000) law for the
wavelength dependence of the attenuation.  Although this law is widely
used in the high-redshift literature, reddening laws are known to
vary, at least in the local universe, due to variations in dust
composition and dust size distribution.  Dust grains can grow to
larger sizes as the density of the gas to which they are mixed
increases, leading to a larger ratio of visual extinction to
reddening $R_V \equiv A_V / E(B-V)$ (Maiolino \& Natta 2002 and
references therein).  At larger lookback times as well, such
variations in environment may cause the extinction curve to vary
between galaxies of different type.  Along these lines, Reddy et
al. (2010) recently argued that young $z \sim 2$ galaxies with ages $<
100$ Myr may follow an extinction curve that is different from
Calzetti, and instead more SMC-like.

Here, we briefly consider how our results would change when adopting
the SMC reddening law from Pr\'{e}vot et al. (1984) and Bouchet et
al. (1985), which increases more steeply towards short wavelengths
compared to the relatively gray Calzetti law.  Naturally, when
adopting the SMC law, less dust (i.e., a smaller $A_V$) is needed to
reproduce the observed UV spectral slope.  Furthermore, it is
interesting to note that without imposing an ad hoc constraint on the
minimum $\tau$, the best-fit $\tau$ and age distribution are less
skewed towards short $\tau$ and young ages.  For the same weak
constraints on $\tau$ ($\log \tau_{min} > 7$), we find that the
underestimate of $SFR_{SED}$ with respect to $SFR_{UV+IR}$ is reduced
to 0 - 0.2 dex at low $SFR_{UV+IR} \lesssim 20\ M_{\sun}/yr$ when
adopting the SMC law.  At high SFRs on the other hand, the discrepancy
becomes worse.  In terms of $\chi ^2$, half of the low $SFR_{UV+IR} <
20\ M_{\sun}/yr$ systems are better fit with a SMC law than a Calzetti
law, while above $SFR_{UV+IR} > 20\ M_{\sun}/yr$ this fraction drops
to 20\%.

From the comparison to $SFR_{UV+IR}$ and the relative distribution of
$\chi^2$ values, we thus conclude that our results are at least
qualitatively consistent with a scenario where steeper extinction
curves are appropriate for lower SFR systems.  Physically, this may
stem from a dust grain distribution that is biased to small sizes in
these environments that are typically of lower gas density.  For
simplicity, consistency with previous work, and because a good SFR
calibration can be obtained provided an appropriate $\tau_{min}$ (see
Section\ \ref{SEDvsUVIR.sec}), we will proceed in this paper to only use the
Calzetti law.  More direct probes of the extinction curve at high
redshift, and its variation with galaxy luminosity, age, and/or SFR
are desired.

\section {SFRs from $H\alpha$}
\label{SFR_Ha.sec}

% FIG 7
%%%%%%%
% FIG 7
%%%%%%%
\begin {figure*}[t]
\centering
\plotone{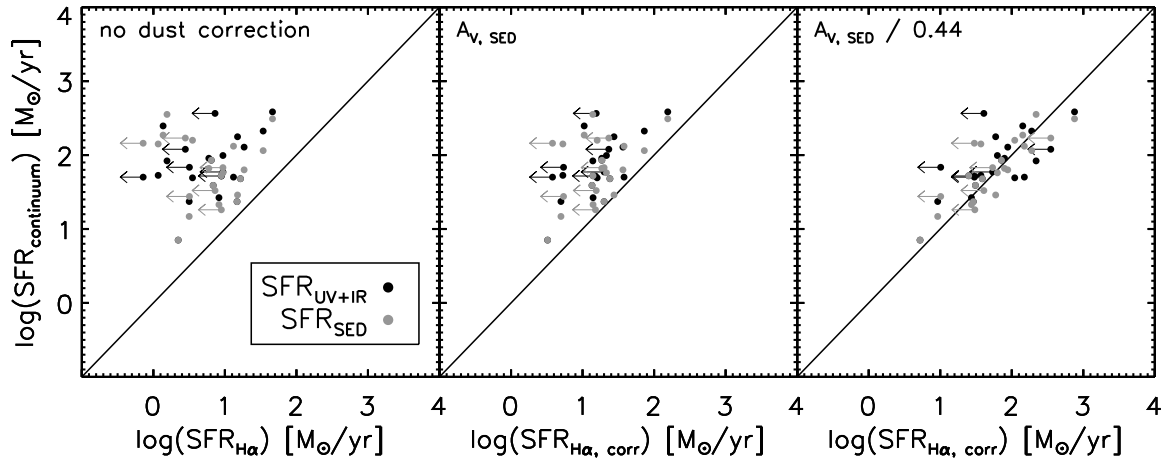} %{figeps/comp_SFRHa.eps}
\caption{SFRs of SINS galaxies at $1.5 < z < 2.6$ as derived from continuum measurements (UV+IR and SED modeling methods) compared to the star formation rate 
as inferred from the $H\alpha$ line luminosity under various assumptions about its extinction.  In the absence of a dust correction, or when simply applying the dust 
correction as inferred from broad-band SED modeling, the $H\alpha$-based SFR is clearly underestimated with respect to other indicators.  Adopting the locally 
calibrated extra attenuation towards HII regions compared to the continuum yields a better correspondence to the other SFR indicators.
\label{SFRHa.fig}}
\end {figure*}

We established self-consistent recipes for SFRs based on UV + IR and
SED modeling.  Now, we extend the cross-calibration of SFR indicators
to measurements of the $H\alpha$ line luminosity, as obtained by
SINFONI on VLT as part of the SINS survey (F\"{o}rster Schreiber et
al. 2009).  The benefit of using integral field spectroscopy is the
absence of slit losses, which complicate the absolute calibration of
total $H\alpha$ luminosities from slit-based measurements (e.g., Erb
et al. 2006; Reddy et al. 2010).  This sample comprises 25
star-forming galaxies in the GOODS-South field spread over the
redshift range $1.5 < z < 2.6$, out which 19 are significantly ($>
3\sigma$) detected in $H\alpha$.  SINFONI exposure times ranged from 1
to 8.5 hours.  Tests on partial data sets of the longest exposed
targets indicate that possible sensitivity-driven flux losses for the
shortest exposed objects are $\sim 30$\%.  19 out of 25 sources are
significantly detected in the MIPS 24 $\mu$m band, and for 9 of those
we extracted PACS fluxes with $S/N > 3$.  We derived monochromatic
$L_{IR}$ using the longest wavelength IR band that has a significant
detection.  In addition to $SFR_{UV+IR}$ values for the IR-detected
sources, we also compute SED modeled SFRs for all 25 galaxies in this
sample.  The two continuum-based estimators are contrasted to the
line-based measurement $SFR_{H\alpha}$ in Figure\ \ref{SFRHa.fig}.

The H$\alpha$ fluxes were converted to star formation rates
$SFR_{H\alpha}$ following the prescription of Kennicutt (1998), which
assumes case B recombination, corrected to a Chabrier (2003) IMF.  Not
unexpectedly, the rates are severely underestimated when no dust
correction is applied ({\it left panel}).  Using the Calzetti et
al. (2000) reddening law, the visual extinction $A_V$ translates to
the extinction at the rest-wavelength of H$\alpha$ as $A_{H\alpha} =
0.82 A_V$.  When adopting the visual extinction as inferred from
broad-band SED modeling $A_{V,\ SED}$ ({\it middle panel}), we find
the offsets in SFR to be reduced, but still amounting to
underestimates of 0.55 dex.

Indeed, it was found by F\"{o}rster Schreiber et al. (2009) that the
SINS data are consistent with roughly twice higher attenuation towards
the HII regions relative to the bulk of the stars.   Pioneering measurements of the Balmer
decrement $H\alpha / H\beta$ in high-z galaxies seem to confirm the
increased extinction towards HII regions, albeit with a significant
galaxy-to-galaxy scatter in the precise degree of differential
extinction (Yoshikawa et al. 2010; Muzzin et al. 2010;
Buschkamp et al. in prep).

In nearby galaxies, it has long been established that HII regions are often
associated with dustier regions than the bulk of the stellar
populations across galaxies (Calzetti et al. 1994, 2000; Cid-Fernandes
et al. 2005).  Studying a sample of low-redshift star-forming and
starburst galaxies, Calzetti et al. (2000) formulated the empirical
calibration $A_{V, neb} = A_{V, SED} / 0.44$ where $A_{V, neb}$ and
$A_{V, SED}$ refer to the visual extinction towards the nebular
regions and the stellar continuum respectively.  

Here, the offset in the middle panel of Figure\ \ref{SFRHa.fig}
reflects this need to take into account differential extinction.  Applying the local
calibration to our high-redshift sample ({\it right panel} of Figure\
\ref{SFRHa.fig}), we find the H$\alpha$-based SFR to line up
well with estimates based on UV+IR and SED modeling.  We
measure a median $\Delta \log SFR = \log SFR_{continuum} - \log
SFR_{H\alpha}$ of 0.08, with a 0.33 dex scatter.  We verified that our
analysis leads to the same conclusion when deriving $A_{V, SED}$ using
M05 models.  Furthermore, the 3 objects in our sample that are X-ray
detected, among which object ID6202 whose broad H$\alpha$ line profile
suggests the presence of AGN activity and/or shocks, do not deviate
from the remaining galaxies in our sample, implying that contributions
to the H$\alpha$ luminosity by non-stellar processes do not dominate
the observed trend.

Using SFR indicators from UV to IR wavelengths, we verified the
findings on differential extinction by F\"{o}rster Schreiber et
al. (2009), which are also consistent with recent work by Onodera et
al. (2010) and Mancini et al. (2011).  In contrast, Reddy et
al. (2010) found that such an extra attenuation correction towards HII
regions may not be necessary.  This may be due to observational
effects (e.g., uncertain slit loss corrections), but differences
between the intrinsic properties of the galaxy samples could play a
role as well.  The galaxies studied by Reddy et al. (2010) typically
have lower stellar masses.  Their extinction may differ due to
different dust properties or distributions.

\section{Life Paths through the SFR versus Mass Diagram: a Consistency Check}
\label{consistency.sec}

With the above calibrated 'ladder of SFR indicators' and SED modeling recipes in
hand, we now perform a galaxy population study, linking progenitors
and descendants in the SFR versus mass diagram over a range of
redshifts, and testing for consistency.  We detail how observed
populations at high redshift are compared to the expected populations
inferred from observations at lower redshift in Section\
\ref{methodology.sec}.  In Section\ \ref{tracingback.sec}, we carry
out this methodology multiple times, varying only the constraints on
the galaxies' SFH.  Here, we make the case for underestimated galaxy
ages as a significant contributor to the observed discrepancies.
After addressing merging as another potential contributor (Section\
\ref{merging.sec}), we return to the issue of constraining galaxy ages
in Section\ \ref{ages.sec}, arguing for an extended wavelength
baseline, or spatially resolved SED modeling as avenues for improvement.

\subsection{Methodology}
\label{methodology.sec}
% SFRs in hand
% its integral, modulo stellar mass loss, the assembled stellar mass
% main sequence of SF, quiescent galaxies
% consistency check
% When IR available, use it as additional constraint in SED modeling
% ignoring merging
% SFHs in SFR vs Mass

If we ignore merging, a galaxy with a mass $M_{obs}$ in stars at the
epoch of observation can be considered as one unit since its onset of
star formation at time $t_{form}$, and built up its stellar content
following

\begin {equation}
M_{obs} = \int_{t_{form}}^{t_{obs}} \! SFR(t - t_{form}) f_{loss}(t_{obs} - t) \, \mathrm{d}t.
\label{continuity.eq}
\end {equation}

Here, the fraction $f_{loss}(t)$ accounts for stellar mass loss.  This
fraction of mass remaining in stars first declines rapidly once the
first supernovae go off, then decreases more gradually by winds from
AGB stars, reaching $\sim 0.6$ after 1 Gyr, and eventually levels off at
$\sim 0.5$ for a Chabrier IMF (BC03).

As age will be an important parameter in our analysis, it is useful to
point out that two definitions of galaxy age are frequently used in
the literature; one being the time since the onset of star formation:
\begin {equation}
age = t_{obs} - t_{form},
\end {equation}
and the other being a measure of the age of the bulk of the stars:
\begin {eqnarray}
age_w = \int_{t_{form}}^{t_{obs}} \! SFR(t - t_{form}) (t_{obs} - t) \, \mathrm{d}t\ / \nonumber \\
\int_{t_{form}}^{t_{obs}} \! SFR(t - t_{form}) \, \mathrm{d}t.
\label{agew.eq}
\end {eqnarray}
For exponentially declining SFHs, $age_w$ ranges between $0.5 \times
age$ ($\tau = \infty$) and $1 \times age$ ($\tau = 0$).  Throughout
this paper, when referring to $age_w$, we will explicitly state
'SFR-weighted age'.

Given equation\ \ref{continuity.eq}, a particular functional form (in
our case parametrized by $\tau$) for the star formation history
$SFR(t)$ translates to a unique path $SFR(M)$ of the galaxy in the SFR
versus mass diagram.  Consequently, the constraints on $M_{obs}$,
$\tau$, and $SFR_{obs}$ allow us to determine the location of each
galaxy's progenitor in the SFR-M diagram at any point
earlier in time.  For galaxies with a PACS and/or MIPS detection, we
apply the IR constraints to the SED modeling by searching for the
minimum $\chi ^2$ solution within a 0.1 dex range around
$SFR_{UV+IR}$.  This approach reduces the degrees of freedom by one.
In addition, it avoids the underestimates by $SFR_{SED}$ at the
high-SFR end (see Section\ \ref{SEDvsUVIR.sec}).  For galaxies that
lack a far- or mid-IR detection, we simply use the SED modeling
performed without constraints on the SFR.

We then compare the galaxy population observed at $1<z<2$ to that
expected at that epoch from tracing the $0<z<1$ population back in
time.  Likewise, we confront the observed $2<z<3$ population to that
backtraced from $1<z<2$, and the $3<z<4$ population to that inferred
from galaxies at $2<z<3$.  Under our premise that galaxies satisfy a
continuity equation over cosmic time, both the location of the main
sequence of star formation, and how densely it is populated should
match.  The presence of any discrepancies could hint at
ill-characterized SFHs, breaking of number conservation by merging, or
cosmic variance leading to one redshift interval being under- or
overdense with respect to the other.

%%%%%%%
% FIG 8
%%%%%%%
\begin {figure*}[htbp]
\centering
%\epsscale{0.8}
%\vspace{0.1in}
\plotone{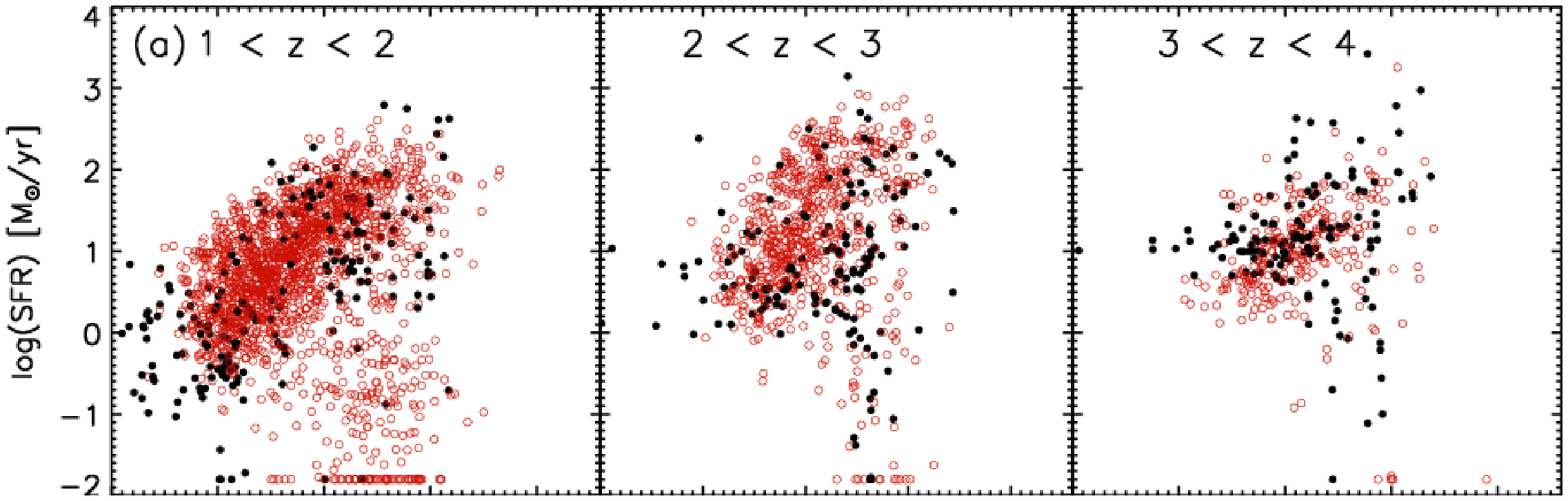} %{figeps/traceback2_goodss_exp_bc03_ch_2D.eps}
%\vspace{0.1in}
\plotone{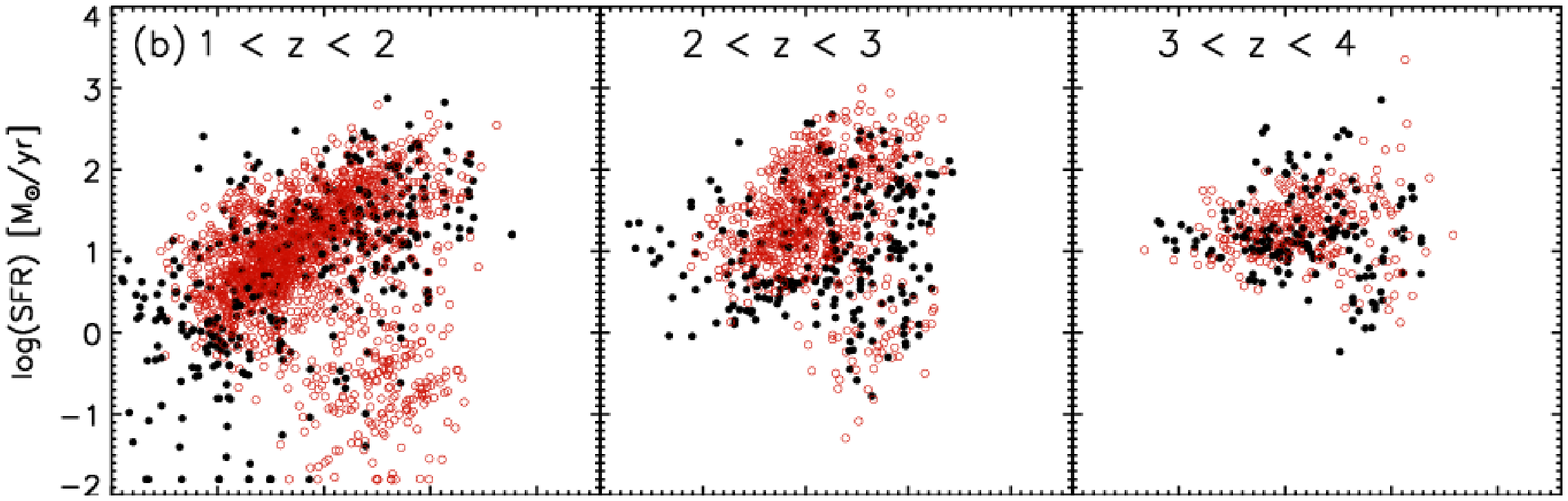} %{figeps/traceback2_goodss_ltaugt8_5_exp_bc03_ch_2D.eps}
%\vspace{0.1in}
\plotone{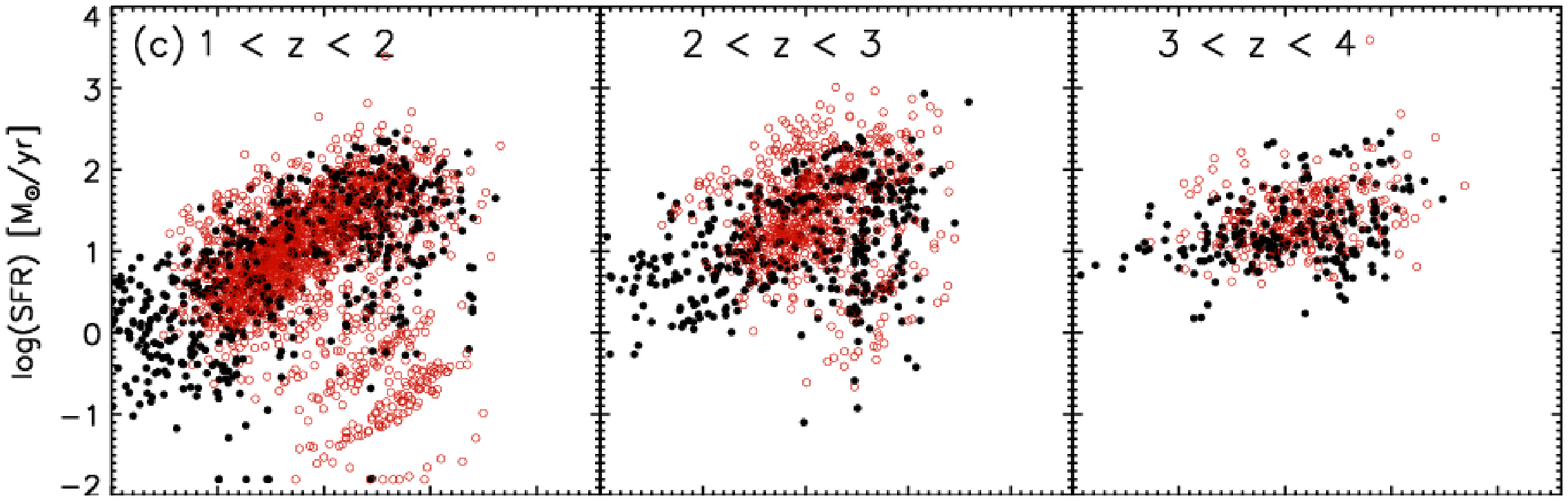} %{figeps/traceback2_goodss_ltaugt8_5_del_bc03_ch_2D.eps}
%\vspace{0.1in}
\plotone{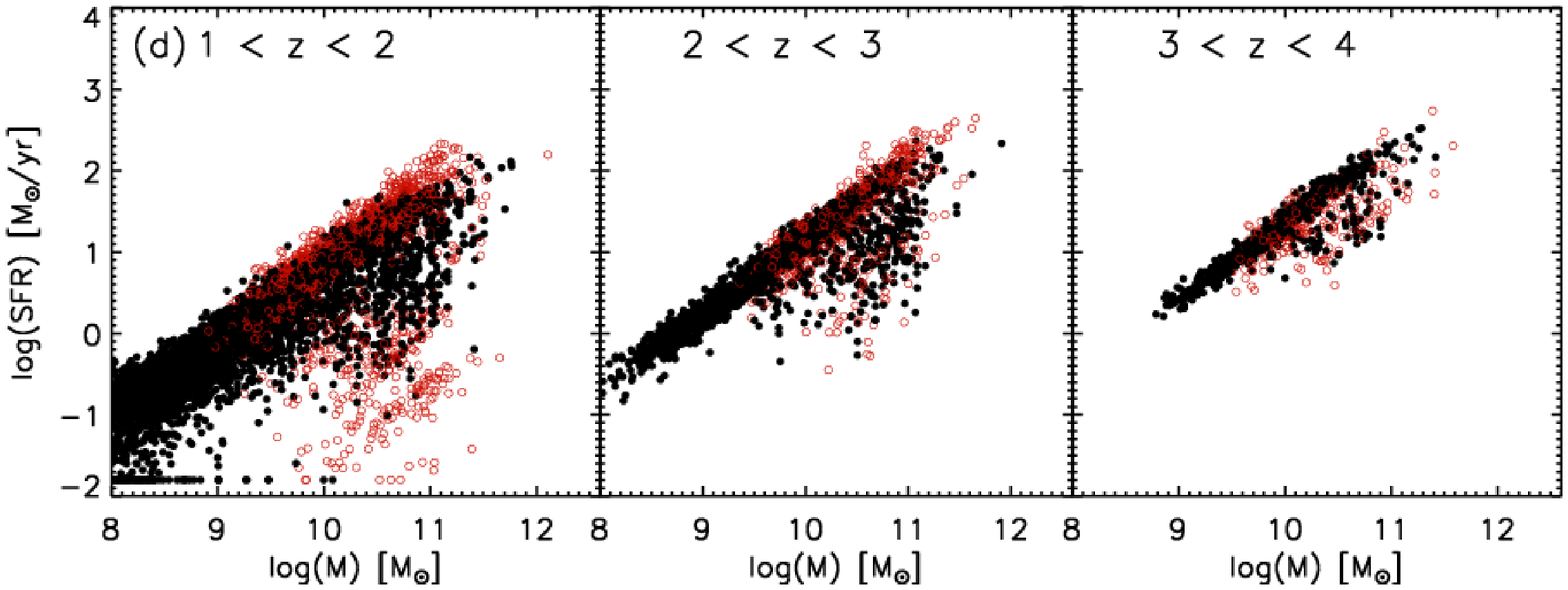} %{figeps/traceback2_goodss_ltaugt8_5_maxold_del_bc03_ch_2D.eps}
%\epsscale{1.0}
\caption{
Observed ({\it red}) SFR versus stellar mass diagram in three redshift bins, compared to the distribution and population expected for the same volume from 
tracing galaxies in the previous redshift bin back in time ({\it black}). The following constraints were applied in modeling the SEDs with BC03 templates: (a) $\tau$ models with $\log (\tau_{min}) = 7.5$, (b) $\tau$ models with $\log (\tau_{min}) = 8.5$, (c) delayed $\tau$ models with $\log (\tau_{min}) = 8.5$, (d) delayed $\tau$ models with $\log (\tau_{min}) = 8.5$ and all galaxies being maximally old.  While occupying a similar region in the SFR versus mass diagram as the observed galaxies, the backtraced population is undernumerous at all redshifts in cases (a) to (c), suggesting an overall underestimate of formation redshifts, especially when no stringent constraints on e-folding time and/or age are imposed.
\label{trace2D.fig}}
\end {figure*} 

\subsection{Tracing Galaxies back in Time}
\label{tracingback.sec}

\subsubsection{$\tau$ Models}
We first perform the above described galaxy population analysis
adopting exponentially declining SFHs with a large range of allowed
e-folding timescales ($\tau > 30$ Myr).  In Figure\
\ref{trace2D.fig}a, we contrast the observed SFR-M relation ({\it
  red}) to that inferred from the SED modeled SFHs of galaxies in the
previous redshift bin ({\it black}).  The black and red symbols were
plotted in a random order, in order to preserve information on the
relative abundance of the two populations, even in crowded regions of
the diagram.  At all redshifts, the backtraced model matches the
observed main sequence of star formation reasonably well in terms of
zeropoint, slope, and scatter.  In particular, even though $\tau$
models by definition do not allow for increasing SFRs, the upwards
shift of the main sequence from $z = 0$ to $z=3$ is captured by the
model.  As is empirically well established, galaxies with lower
specific star formation rates exist alongside normal star-forming
galaxies at least out to $z \sim 3$ (see, e.g., Labb\'{e} et al. 2005;
Kriek et al. 2006).  Their relative abundance is an increasing
function of galaxy mass, and declines with increasing redshift
(Fontana et al. 2009).  We note that, while these quiescent systems
are abundantly present at $1<z<2$, they are missing from the
backtraced population computed for that redshift bin.  Galaxies with
similarly low specific star formation are present in our $0<z<1$
sample.  This suggests that we are not merely missing the descendants
of $1<z<2$ quiescent galaxies due to the smaller volume probed at
$0<z<1$.  In general, it is the case that the backtraced number
densities, both on and off the main sequence, are at all redshifts
lower than those directly observed.

\begin {figure*}[htbp]
\centering
\plottwo{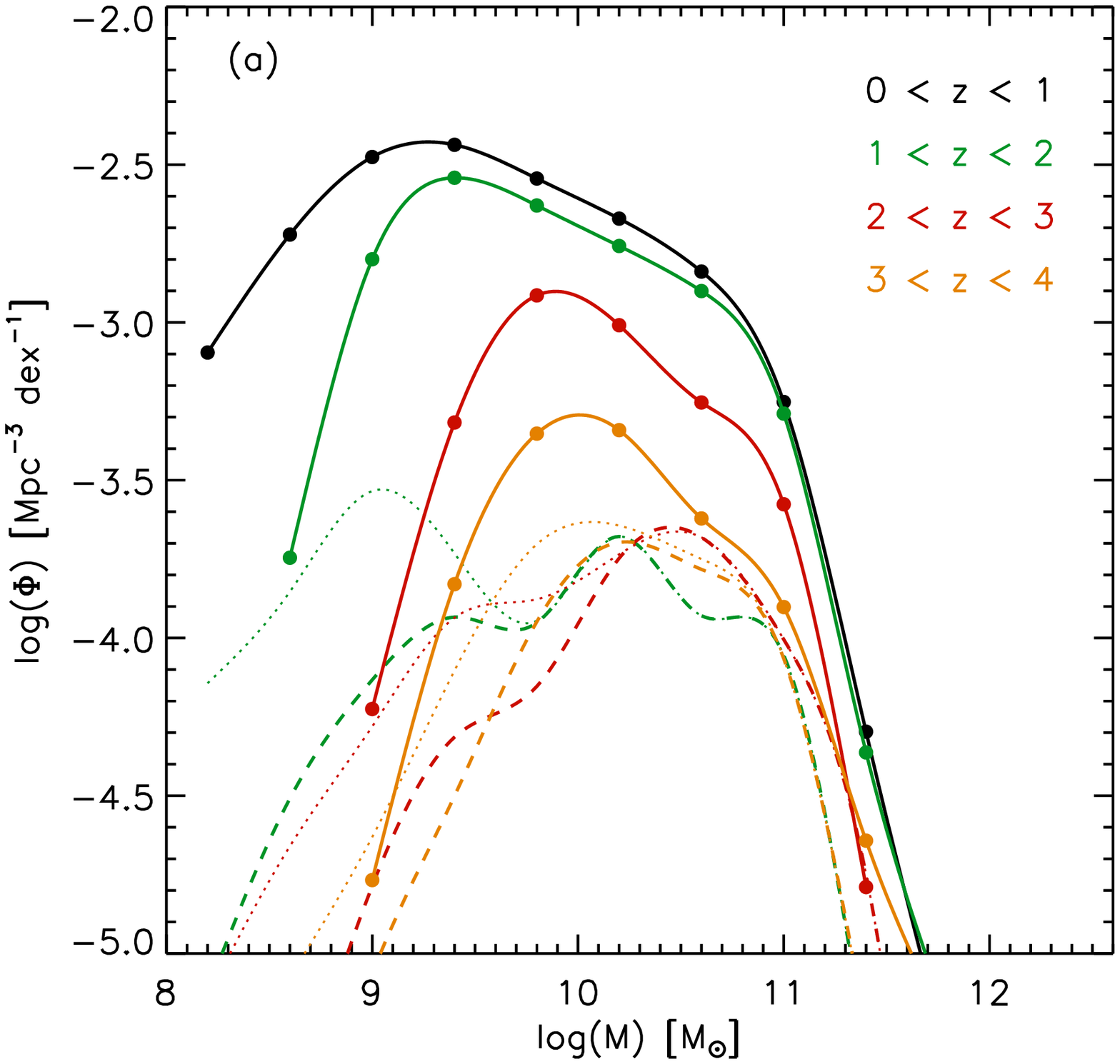}{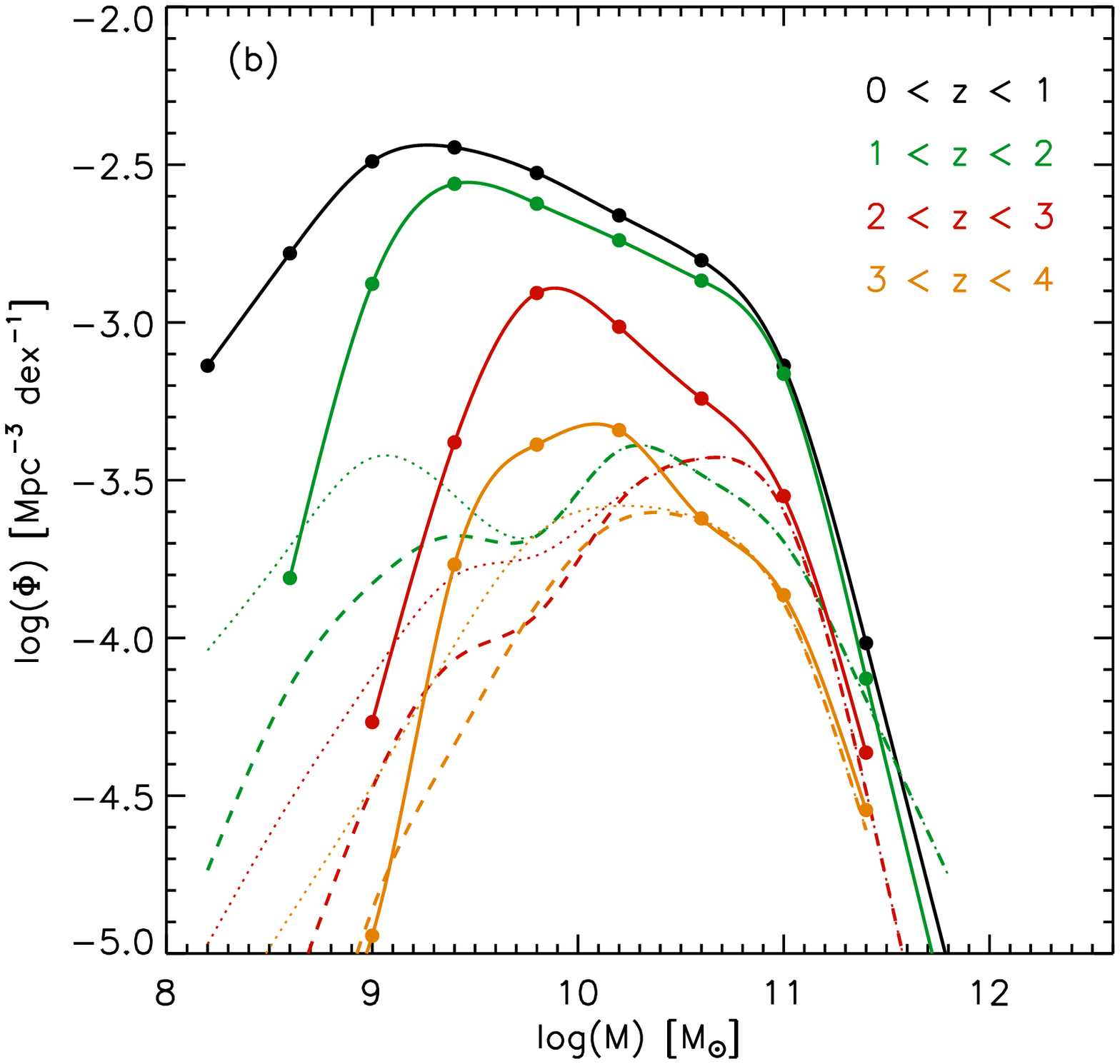} %{figeps/traceback2_goodss_exp_bc03_ch_1D.eps}{figeps/traceback2_goodss_ltaugt8_5_exp_bc03_ch_1D.eps}
\plottwo{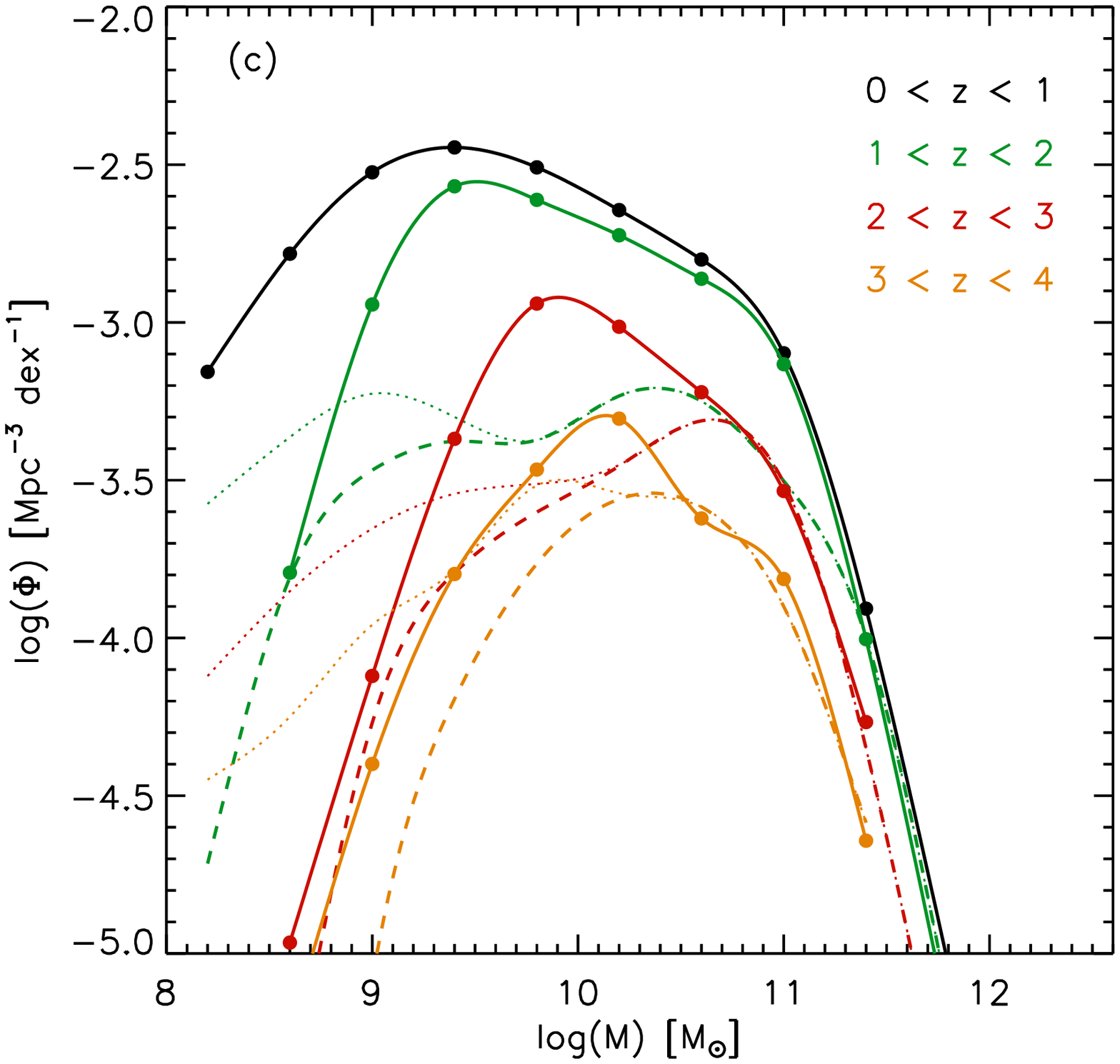}{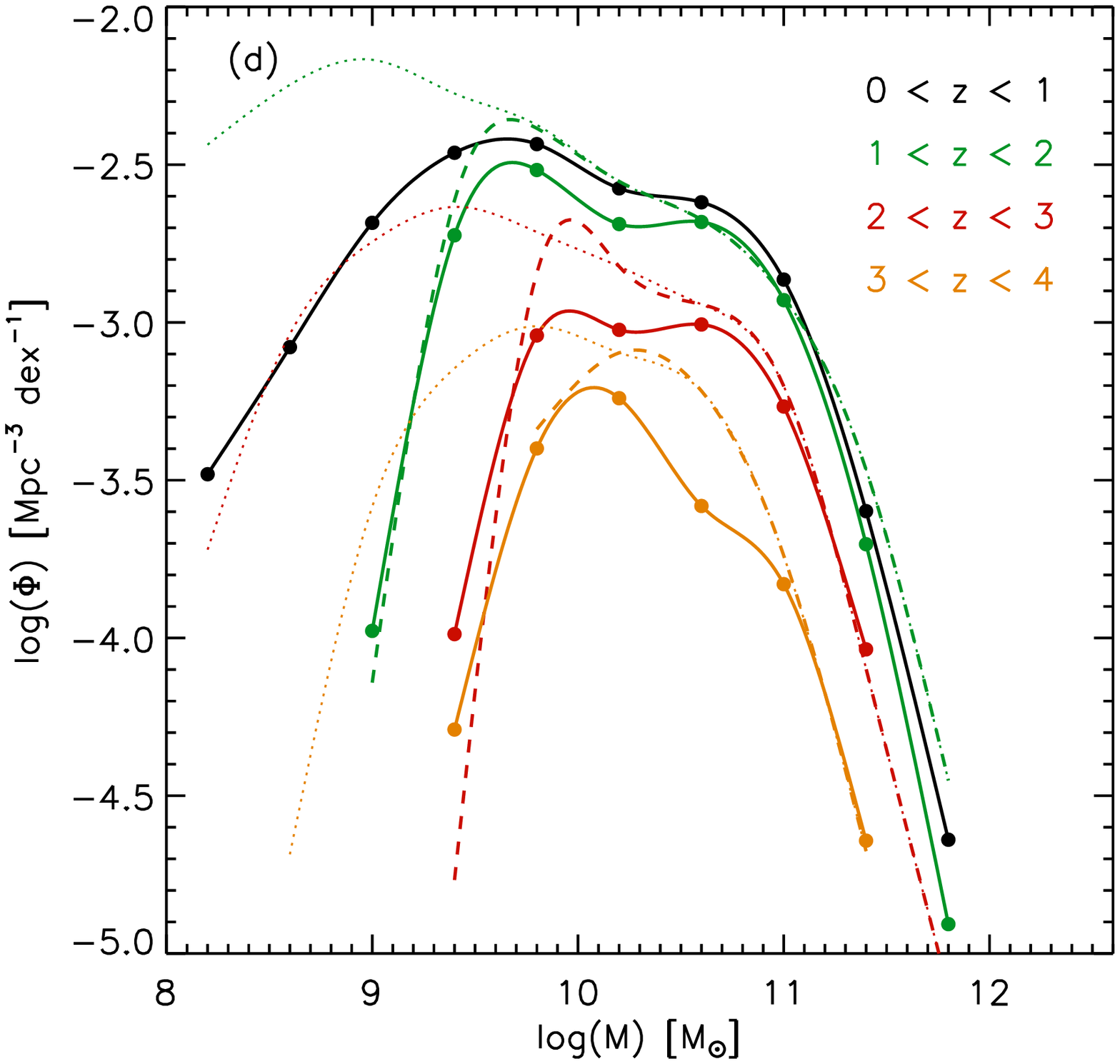} %{figeps/traceback2_goodss_ltaugt8_5_del_bc03_ch_1D.eps}{figeps/traceback2_goodss_ltaugt8_5_maxold_del_bc03_ch_1D.eps}
\caption{Mass functions from $z=0$ to 4 as observed ({\it solid curves}), and as obtained by tracing lower-redshift galaxies back in time using the constraints on their 
SFHs from SED modeling ({\it dotted curves}).  No incompleteness corrections are applied.  Instead, dashed lines show the mass function of the traced-back 
populations after applying the $K = 24.3$ magnitude limit.  The following constraints were applied in modeling the SEDs with BC03 templates: (a) $\tau$ models with 
$\log (\tau_{min}) = 7.5$, (b) $\tau$ models with $\log (\tau_{min}) = 8.5$, (c) delayed $\tau$ models with $\log (\tau_{min}) = 8.5$, (d) delayed $\tau$ models with $\log 
(\tau_{min}) = 8.5$ and all galaxies being maximally old.  Chi-square minimization prefers to fit steeply declining SFHs with young ages.  Such solutions are physically 
unrealistic, as they violate the galaxy continuity equation as tested here.  Instead, star formation over more prolonged timescales seems preferred.
\label{trace1D.fig}}
\end {figure*}

This result is quantitatively better appreciated when we collapse the
SFR-M diagram along the SFR axis, resulting in the mass functions
plotted in Figure\ \ref{trace1D.fig}a.  Here, the solid lines
represent the observed mass function at different redshifts.  In
agreement with Marchesini et al. (2009), whose analysis was partly
based on the same data set, we observe fairly little evolution in
number density at the high-mass ($ > 10^{11}\ M_{\sun}$) end, compared
to a steeper build-up of stellar mass with cosmic time in the
intermediate ($\sim 10^{10}\ M_{\sun}$) regime.  We did not correct
the observed mass functions for incompleteness.  Instead, we apply the
$K = 24.3$ observed magnitude limit of the FIREWORKS catalog to the
galaxies that were evolved back in time, leading to the dashed curves
in Figure\ \ref{trace1D.fig}a.  This approach allows for a fair
comparison as long as galaxies' higher-z progenitors have fainter
observed $K$-band magnitudes than their descendants at the epoch of
observation.

Figure\ \ref{trace1D.fig}a clearly shows that the backtraced
population, as based on SED modeling with $\tau$ models with a large
freedom in $\tau$, is underpopulated.  The underestimate of the number
densities increases with decreasing stellar mass, becoming dramatic at
intermediate masses, particularly for the lower redshift bins, where
it exceeds an order of magnitude.  As we will see in Section\
\ref{merging.sec}, such large differences in number density are
unlikely to result from merging of galaxies alone, a process that we
ignored so far in our analysis.  An underdensity of the GOODS-South
field in the $0<z<1$ interval (Wolf et al. 2003) may contribute to the
offset seen between the observed and backtraced $1<z<2$ population.
However, the same trend is seen at all redshifts.  This cannot be due
solely to field-to-field variations unless each redshift bin is
underdense relative to the one above.  Moreover, the magnitude of the
offsets seems incompatible with field-to-field variations.  Following
the methodology of Somerville et al. (2004), the expected uncertainty
due to cosmic variance on the number density of $\log M > 10.5$
galaxies at $0<z<1$, $1<z<2$, $2<z<3$, and $3<z<4$ as derived from the
GOODS-South field is 34, 24, 26 and 32\% respectively, much smaller
than the discrepancies between the observed and backtraced populations
in Figure\ \ref{trace1D.fig}a.  The most straightforward
interpretation is therefore that the SED modeled stellar populations
are biased towards young stellar ages.  Given an underestimated age,
we would infer the stellar mass of a galaxy to be built up relatively
recently.  Or in other words, as we trace this galaxy back in time,
its stellar mass would decrease too rapidly, leading to a lower
formation redshift at which the galaxy drops out of the backtraced
sample.  The origin of such age underestimates lies most plausibly in
the fact that younger stellar populations outshine any underlying old
population in the integrated SED, especially at the shorter
wavelengths.  Because of this outshining effect, any deviation from
the simple SFH and foreground dust distribution assumed makes it
impossible to correctly recover the true SFR-weighted age (Wuyts et
al. 2009a; Maraston et al. 2010).

Somewhat earlier formation epochs are obtained when requiring longer
e-folding timescales of $\tau > 300$ Myr.  Since this also provided a
better agreement between SFR indicators (Section\
\ref{SEDvsUVIR.sec}), we adopt this value of $\tau_{min}$ in the
remainder of our analysis.  In the median, the SFR-weighted age of the
galaxies in our sample then increases by 0.15 dex.  The
high-mass end of the observed and backtraced mass functions now show a
good agreement (Figure\ \ref{trace1D.fig}b).  At lower masses, the
offsets remain present, although reduced by a few 0.1 dex.  The mass
dependence of the offsets may imply that age underestimates are more
of a concern for low- to intermediate-mass galaxies than for the most
massive (often older) systems.

%%%%%%%
% FIG 9
%%%%%%%

\subsubsection{Delayed $\tau$ Models}
\label{delayedtau.sec}
Recently, several authors have argued, both on observational (Renzini
et al. 2009; Maraston et al. 2010; Papovich et al. 2011) and on
theoretical (Finlator et al. 2007; Lee et al. 2010) grounds, that
galaxies may undergo increasing, rather than decreasing, SFHs during
parts of their life, specifically at early times ($z \gtrsim 3$).  To
allow for such phases of increasing SFR without imposing an artificial
breakdown of our galaxy sample in sources fitted with strictly
increasing versus sources fitted with strictly decreasing SFHs, we
repeated our analysis using the following SFH:

\begin {equation}
SFR(t) = t e^{-t/\tau}
\end {equation}

These so-called delayed $\tau$ models allow for solutions with
increasing SFRs (for ages $< \tau$) as well as solutions in which the
SFR is decreasing after a prior phase during which it was increasing
(for ages $> \tau$).  At $t \ll \tau$, $SFR(M) \sim \sqrt{M}$ for
delayed $\tau$ models as opposed to $SFR(M) \approx constant$ for
simple $\tau$ models.  This corresponds to evolutionary tracks with
steeper initial slopes in the SFR-M diagram than default $\tau$
models, but still shallower than many literature values for the slope
of the main sequence.  The latter vary between this slope of a half
and a slope of unity.  Its precise value is still widely debated
(Noeske et al. 2007; Cowie \& Barger 2008; Pannella et al. 2009;
Santini et al. 2009; Rodighiero et al. 2010), and may evolve with
redshift (Dunne et al. 2009).

Figure\ \ref{agetau.fig}a shows the $age / \tau$ ratio obtained by
fitting delayed $\tau$ models to the SEDs of $0<z<4$ galaxies as
function of SFR.  Overall, only 20 \% of the galaxies is best
characterized by being observed during a phase in which the SFR is
increasing ($age < \tau$).  However, this fraction is a strong
function of the level of star-forming activity (and therefore
redshift).  At SFRs above 10, 50, and 100 $M_{\sun}/yr$, galaxies that
are best fit by rising SFRs account for 40, 60, and 70\% of the
overall population.  At $z > 2$, 37\% of the galaxies in our $K <
24.3$ sample are found to be in the rising part of their SFH (i.e.,
$age < \tau$).  We note that the reduced chi-square ($\chi^2_{red}$)
values of the fits with delayed $\tau$ models are indistinguishable
from those obtained by fitting simple $\tau$ models.  The SFR-weighted
ages are in the median 0.1 dex older than those inferred from simple
$\tau$ models, with a scatter of 0.15 dex, and a weak trend with age
(the systematic offset being smaller for systems that formed the bulk
of their stars more than a Gyr ago).  Despite the modest increase in
estimated ages, it is clear from Figure\ \ref{trace2D.fig}c and\
\ref{trace1D.fig}c that the underestimate of galaxy number densities at
intermediate masses remains significant for this alternative SFH.  In
other words, in the absence of any stringent constraints on age, the
delayed $\tau$ models are subject to the same outshining effect as
were the simple $\tau$ models.

\subsubsection{Maximally Early Onset of Star Formation}
\begin {figure}[htbp]
\centering
%\epsscale{0.5}
\plotone{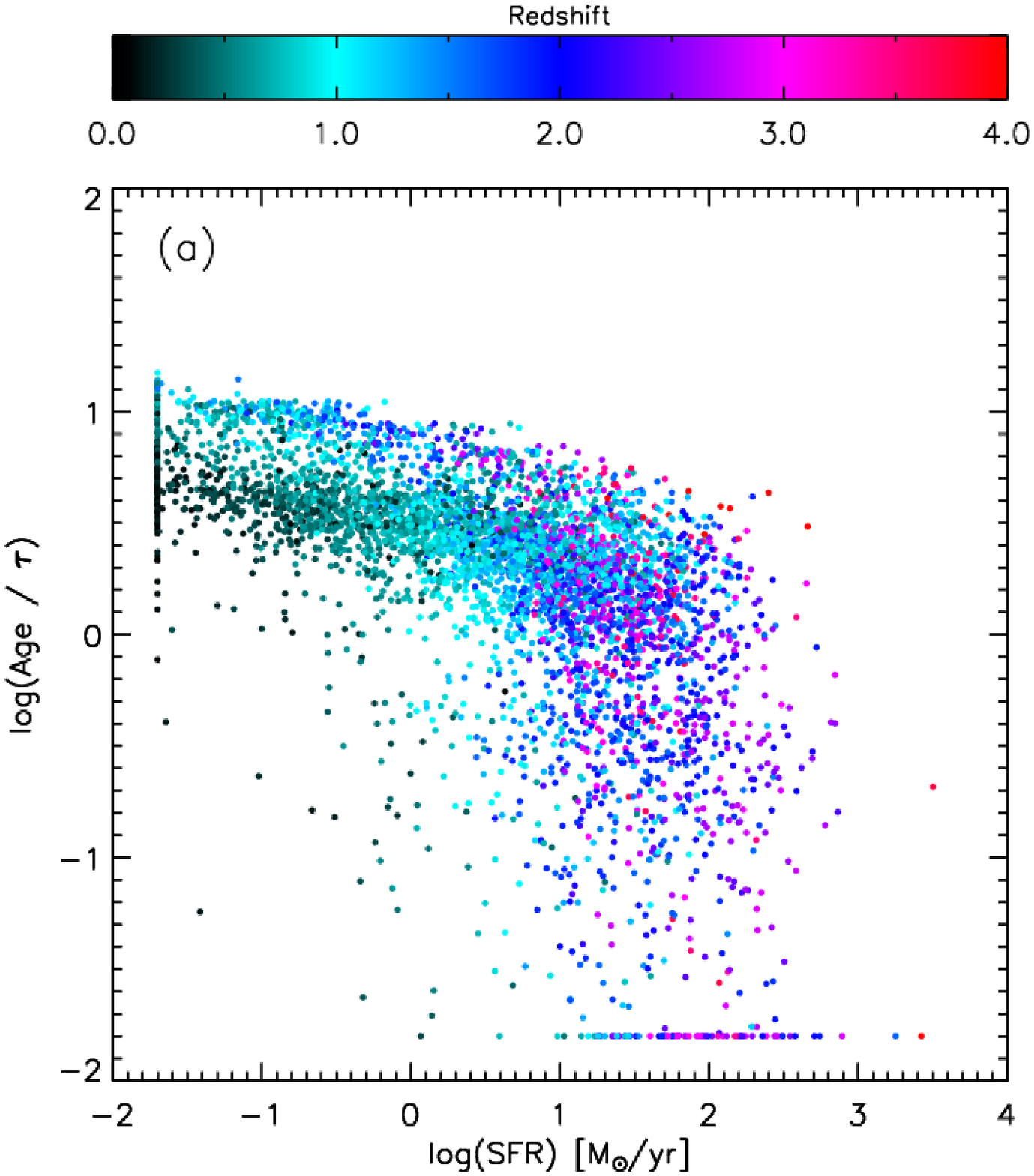} %{figeps/a2t_ltaugt8_5_del_bc03.eps}
\plotone{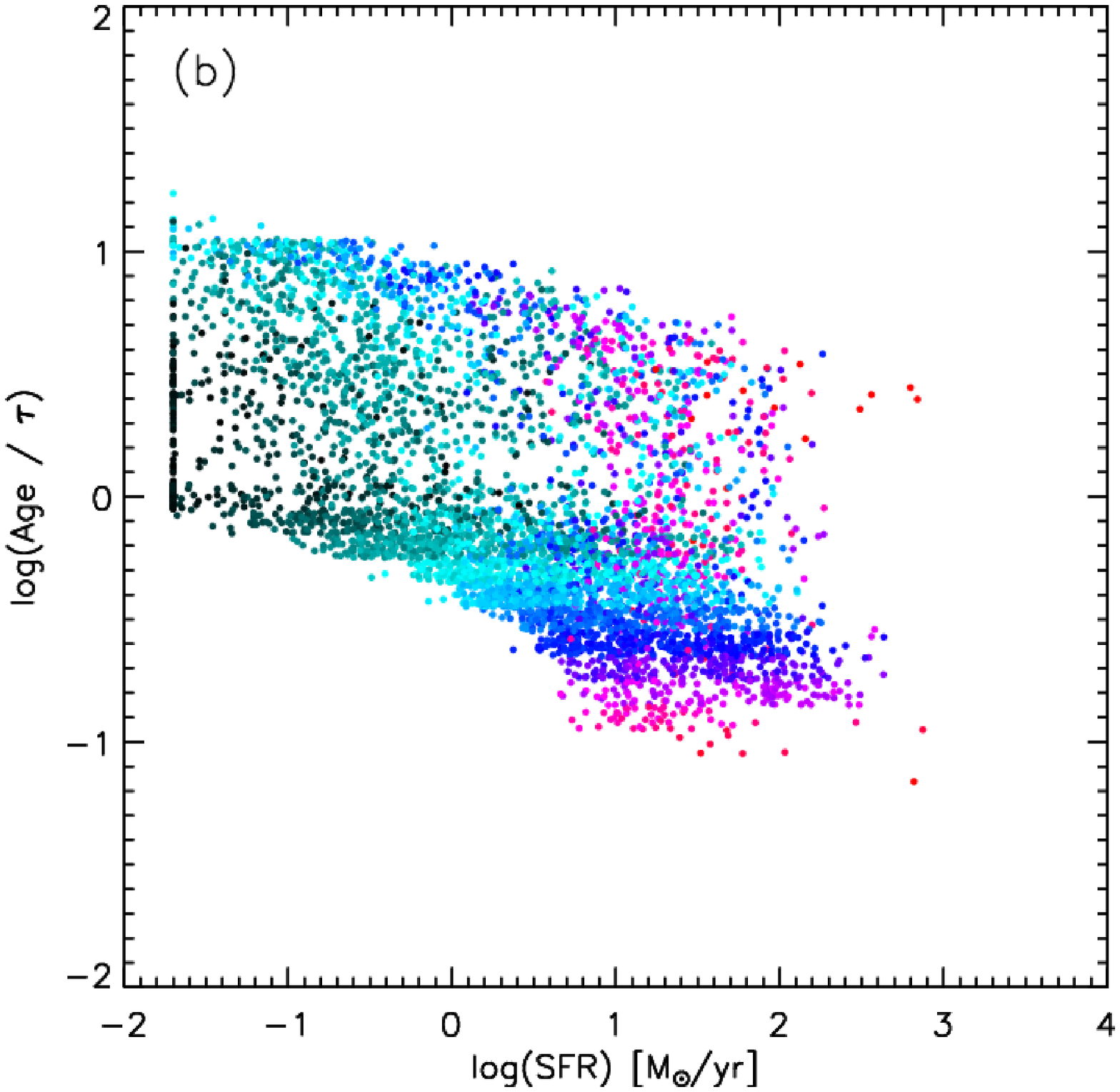} %{figeps/a2t_ltaugt8_5_maxold_del_bc03.eps}
%\epsscale{1.0}
\caption{Ratio of age (i.e., time since onset of star formation) over $\tau$ as function SFR for delayed $\tau$ models with (a) weak constraints on age (age $> 50$ Myr), and (b) maximally old ages.  When leaving age free, only a fraction of galaxies at the high SFR end is best modeled as following a rising SFH at the epoch of observation.  When imposing very high formation redshifts ($z_{form} \sim 10$), larger values of $\tau$ ($>$ age) are preferred.
\label{agetau.fig}}
\end {figure}

Finally, we repeat the population analysis once more, now making the
assumption that all galaxies are maximally old (i.e., $age =
t_{Hubble}(z_{obs})$).  The SFR-weighted age $age_w$ may (and will)
still be younger than the age of the universe at the epoch of
observation, but we perform the SED modeling forcing the onset of star
formation to take place at very high redshift.  Given the discreteness
in age of stellar population models, small variations in the adopted
$z_{form}$ occur depending on the galaxy's precise redshift, but it is
typically around $z_{form} \sim 10$ and the precise value is
irrelevant for this exercise.  We again adopt delayed $\tau$ models,
as they result in $\chi ^2_{red}$ values that are better by a factor
of $\sim 1.4$ than those obtained from fitting simple $\tau$ models
with the same restriction on $z_{form}$.  In practice, the least
squares solution is now often one with $\tau > age$ (see the abundance
of galaxies with $\log(age/\tau) < 0$ in Figure\ \ref{agetau.fig}b),
such that the SFR-weighted age is minimized within the allowed range
that goes from 0.5 (for $\tau \approx \infty$) to 1 (for $\tau \approx
0$) times the age of the universe at the observed epoch.  These long
e-folding times, obtained when maximizing the formation redshift,
serve not to violate the constraints from the longer wavelengths on
stellar mass, which tends to be the most robust parameter in SED
modeling (e.g., F\"{o}rster Schreiber et al. 2004; Shapley et
al. 2005).  Nevertheless, we do note that the median stellar mass of
galaxies in our sample increases by 0.25 dex compared to SED modeling
where no maximum formation redshift was imposed.

Although extreme, the assumption of maximally old galaxies is useful
as a sanity check.  If the backtraced galaxy samples are still
underpopulated with respect to the observed samples, the origin of the
discrepancy must most likely be attributed to other effects such as
merging, cosmic variance, or more extreme SFHs in which an even larger
fraction of the stellar mass was assembled at very early times.
Figure\ \ref{trace1D.fig}d shows that this is not the case.  At all
redshifts, the backtraced mass functions match the observed mass
functions surprisingly well, and are if anything slightly
overproducing the number densities in some mass bins (an effect that
is somewhat more pronounced if we had adopted simple $\tau$ models,
that lead to even older SFR-weighted ages).  We note that the SFR-M
relation (Figure\ \ref{trace2D.fig}d) has tightened significantly by
making the rigorous assumption on early formation epochs.  This
indicates that, aside from an intrinsic scatter owing to
galaxy-to-galaxy differences in SFH, the scatter in the observed SFR-M
relation also depends on model assumptions.  Likewise, we find that
conclusions on the differential evolution of the stellar mass function
also depend on the assumed SFH.  When making similar assumptions about
the SFH as Marchesini et al. (2009), we confirm their finding of
little evolution at the high-mass end, and much more at lower masses
(see Figure\ \ref{trace1D.fig}b).  When instead using delayed tau
models in combination with tight constraints on the formation
redshifts, the degree of such differential evolution reduces
significantly (see Figure\ \ref{trace1D.fig}d).

%%%%%%%
% FIG 10
%%%%%%%

\subsection{The Effect of Merging}
\label{merging.sec}

%%%%%%%
% FIG 11
%%%%%%%
\begin {figure}[htbp]
\centering
\plotone{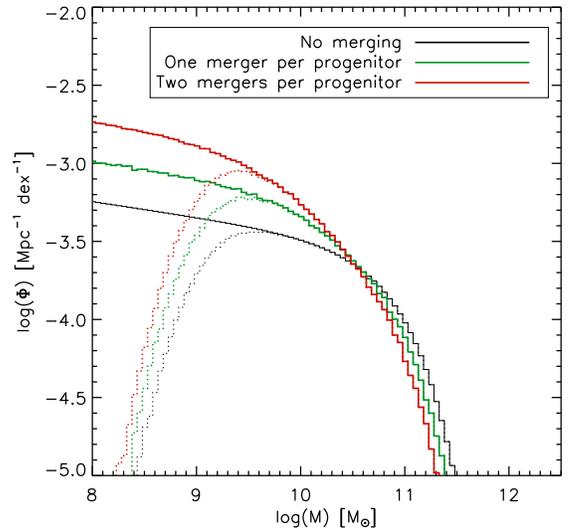} %{figeps/illustrate_merging.eps}
\caption{Toy model exploring the impact of merging on tests of the galaxy continuity equation.  Let the black curve be the high-redshift mass function as constructed by 
evolving a lower redshift observed galaxy population back in time, in
the absence of merging.  In a scenario where each galaxy underwent one merger between the low and 
high redshift bin, with mass ratio chosen randomly between 10:1 and
2:1, the green curve would be the resulting high-redshift mass
function.  Two merger events ({\it red}) or more per galaxy would steepen the mass function further.
\label{merging.fig}}
\end {figure}

So far, we assumed that each observed galaxy had only one progenitor
that assembled its stellar mass by in situ star formation.  In a
hierarchical universe, where dark matter haloes merge to form
increasingly larger structures, carrying with them the baryons that
assembled at the centers of their potential wells, our simplified toy
model must therefore fail to capture the detailed evolution of galaxy
populations at some level.  In order to properly account for galaxies
merging (or in fact de-merging as we trace them back in time), an
accurate knowledge of the merger rate as function of redshift, mass,
and mass ratio would be required.  Moreover, the probability that a
galaxy observed at redshift $z_{obs}$ underwent a merger since
redshift $z_{obs} + \Delta z$ (and hence has to be split in two as we
trace it back to that epoch) may not only depend on its mass.  For
example, if merging leads to quenching (e.g., Di Matteo et al. 2005),
a quiescent galaxy may be more likely to have undergone a merger in
its past than a star-forming galaxy of similar mass.  Furthermore, it
is well motivated that mergers between gas-rich systems trigger
starbursts in their centers, which inhibits evolving galaxies back
along smooth tracks in the SFR-M diagram.  In fact, the scatter in the
SFR-M relation of star-forming galaxies may well be produced by such
burstiness, triggered by (minor) mergers or other processes.  Clearly,
implementing all of these processes in our consistency check is beyond
the scope of this paper.  We therefore choose to illustrate the effect
of merging activity on our analysis by two models of increasing
complexity.

\subsubsection{A Simple Toy Model for Merging}
The first is an idealized toy model, presented in Figure\
\ref{merging.fig}.  Here, the black curve represents a hypothetical
mass function at $z_{obs} + \Delta z$ as traced back from galaxies
observed at $z_{obs}$.  The dotted line accounts for the magnitude
limit of our catalog.  If every galaxy observed at $z_{obs}$ underwent
one merger in the interval $\Delta z$ (for simplicity we assume one
that did not affect the SFH), the actual number of progenitors will be
twice as large as anticipated in the absence of merging.  Since the
mass distribution of progenitors will be shifted to lower values, the
number of galaxies entering our magnitude-limited catalog will be
larger by a somewhat smaller factor.  Under the assumption of a merger
rate that is a constant function of the mass ratio over the range 10:1
to 2:1, the resulting mass function is plotted in green.  If all of
these merger progenitors themselves descended from mergers taking
place since $z_{obs} + \Delta z$, the red curve would be the
backtraced population (consisting of four times as many galaxies as
observed at $z_{obs}$, ignoring incompleteness issues).  The net
effect is a steepening of the mass function, the magnitude of which
depends on the amplitude of the merger rate.

%%%%%%%
% FIG 12
%%%%%%%
\begin {figure}[t]
\centering
\plotone{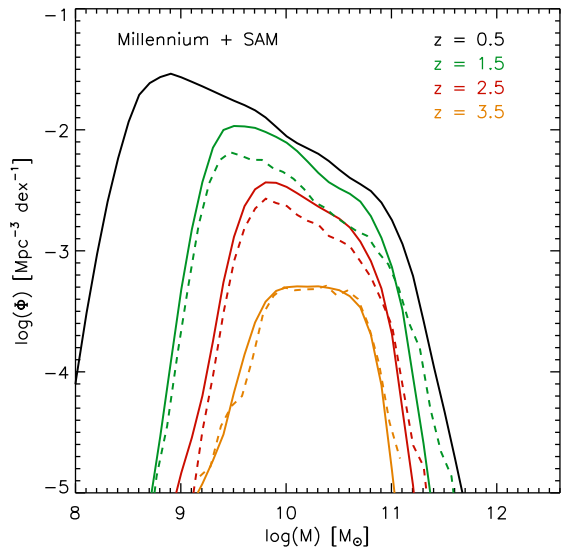} %{figeps/plot_MF.eps}
\caption{
Stellar mass functions down to an observed $K$-band magnitude of 24.3
derived from the De Lucia \& Blaizot (2007) SAM which is based on the
Millennium Simulation ({\it solid lines}).  Dashed lines indicate the
mass functions at $z = 1.5$, 2.5 and 3.5, constructed after summing
those galaxies that have a common descendant at $z=0.5$, 1.5 and 2.5
respectively.  Accounting for merging as we track galaxies back in
time ({\it solid lines}) steepens the mass function with respect to an
approach that ignores merging ({\it dashed lines}).
\label{SAM.fig}}
\end {figure}

\subsubsection{Merging in Semi-Analytic Models}
While conceptually demonstrating the impact of merging, our toy model
does not contain information on the frequency of galaxy collisions,
and therefore its conclusions remain qualitative at best.  In order to
obtain a more quantitative estimate, we turn to a second, more
physically motivated model in Figure\ \ref{SAM.fig}.  Here, the solid
lines mark the galaxy stellar mass functions at a range of redshifts as
extracted from the Munich semi-analytic model (SAM) of galaxy
formation by De Lucia \& Blaizot (2007).  The SAM is rooted in the
pure dark matter Millennium Simulation (Springel et al. 2005), and
hence incorporates the hierarchical mass build-up characteristic for a
$\Lambda$-CDM cosmology.  It furthermore contains a set of
prescriptions to model the baryonic physics within this framework, and
outputs among other parameters the observed-frame photometry, which we
use to apply the $K$-band magnitude limit of the observations.  Here,
we will not focus on differences between the model and observed mass
functions (most notably a relatively late build-up of the high-mass
end and early build-up of the low-mass end in the SAM), these are
discussed at length by Marchesini et al. (2009).  Instead, we use the
SAM with its physically motivated merger rates as a self-consistent
testbed to constrain the impact of merging activity.

Since the merger tree of each galaxy is stored, it is trivial to
perform our analysis from Section\ \ref{methodology.sec} on the SAM,
i.e., to evolve galaxies back in time according to their SFH ignoring
merging.  For example, to obtain the inferred population at $z=2.5$ by
winding back the clock from $z=1.5$, we simply group all $z=2.5$
galaxies that have a common descendant at $z=1.5$, sum their masses,
and construct the mass function from the resulting sample ({\it red
  dashed line in Figure\ \ref{SAM.fig}}).  In the SAM, by
construction, any difference between the backtraced and actual mass
function at a given redshift is due to merging.  As expected from our
simplistic toy model (Figure\ \ref{merging.fig}), the net effect of
including merging is a steepening of the mass function: while the
high-mass end is reduced, the number density at intermediate masses
increases by 0.25 dex, 0.15 dex and 0 dex at $z=1.5$, 2.5 and 3.5
respectively.

We conclude that accounting for mergers will reduce the discrepancy in
number densities seen in Section\ \ref{tracingback.sec} at low to
intermediate masses, but typical merger rates from dark matter +
semi-analytic models seem insufficient to fully resolve the
underdensities we found.  Invoking more active merger histories would
break down the good agreement at the high-mass end ($M \gtrsim
10^{11}\ M_{\sun}$) by suppressing the abundance of those galaxies.
Given these concerns, we now turn to what is likely a major
contributor to the offsets between the observed and backtraced mass
functions in Figure\ \ref{trace1D.fig}a-c: biases in
characterizing the galaxy-averaged stellar age.

\subsection{Constraining Galaxy Ages}
\label{ages.sec}

In Section\ \ref{tracingback.sec}, we found that the backtraced and
observed mass functions over a wide range of redshifts matched better
when forcing maximally old ages in the SED modeling.  Although merging and
stochastic bursts will undoubtly affect the evolutionary tracks of
galaxies in the SFR-M diagram, it seems unlikely that they can account
for the discrepancies observed when leaving formation redshifts as a
free parameter in the SED modeling.  We therefore turn to the
measurement of galaxy ages, and discuss two avenues for improvement.

\subsubsection{Extending the Wavelength Range}
\begin {figure}[t]
\centering
%\epsscale{0.88}
\plotone{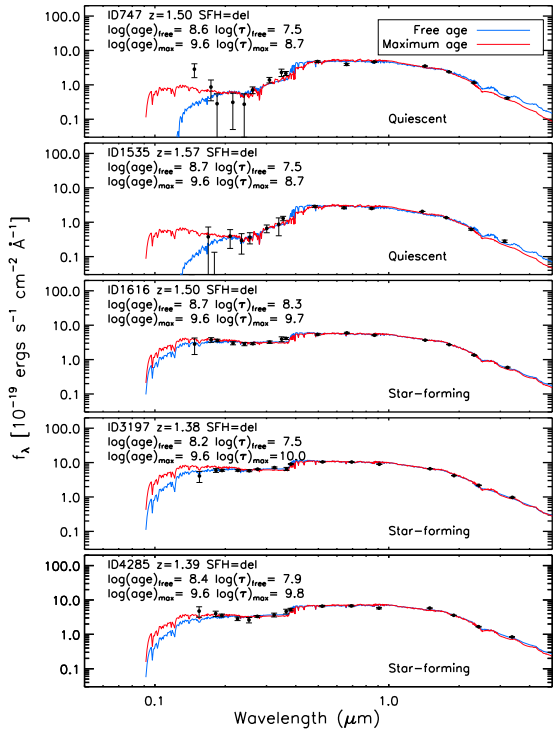} %{figeps/free_vs_maxold_SED.eps}
%\epsscale{1.0}
\caption{Rest-frame broad-band SEDs of $z \sim 1.5$ galaxies, with best-fit BC03 template leaving age free ({\it blue}), or forcing a maximally old onset of star formation ({\it red}).  
The top two panels present quiescent galaxies, whereas the bottom three panels present SEDs of galaxies on the main sequence of star formation.  Stronger 
observational constraints on the rest-frame 0.1 - 0.2 $\mu$m regime are desired to improve the estimate of stellar ages.
\label{free_vs_maxold.fig}}
\end {figure}

First, it is illustrative to contrast the best-fit stellar population
models with and without the constraint of a maximally early onset of
star formation.  The top two panels in Figure\
\ref{free_vs_maxold.fig} show the SEDs of two quiescent galaxies
at $z \sim 1.5$, while the bottom three panels represent cases at the
same redshift of galaxies populating the main sequence of star
formation.  Irrespective of galaxy type, we note that both the free
age and the maximally old best-fit model reproduces the broad-band
SEDs remarkably well.  By and large, the two templates are
indistinguishable for broad-band tracers, except at rest-frame
wavelengths shortward of 0.2 $\mu$m, and to a lesser extent at the
longest wavelengths probed ($\lambda_{rest} \gtrsim 2.5 \mu$m).
Present-day stellar population synthesis codes are facing the largest
differences in the latter regime (Maraston 2005).  We therefore conclude that more stringent
constraints on the $\lambda_{rest} < 0.2 \mu$m SED, from deep FUV,
NUV, and $U$-band observations, would help most to better constrain galaxy
stellar ages without the need to impose ad hoc constraints on
formation redshifts.  Rettura et al. (2010), among others, recognize
this, and succesfully exploit rest-frame UV data of $z \sim 1.2$
ellipticals in clusters and field environments to constrain both their
onset and timescale of star formation.

%%%%%%%
% FIG 13
%%%%%%%

\subsubsection{Resolved SED Modeling}
\begin {figure}[t]
\centering
\plotone{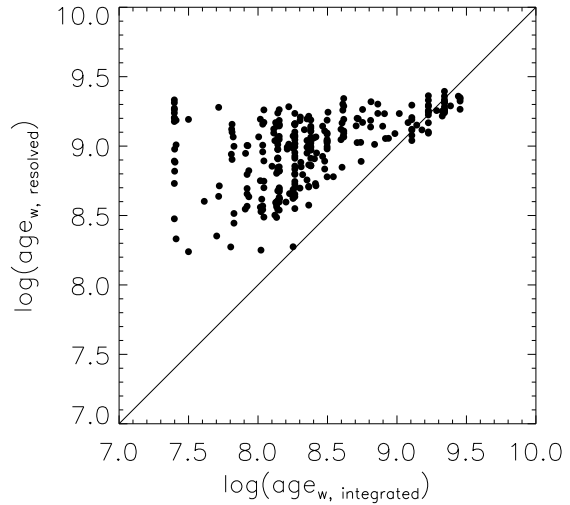} %{figeps/comp_res_agew.eps}
\caption{Comparison of the SFR-weighted age of galaxies with $K <
  24.3$ at $1.5<z<2.5$, as determined by fitting simple $\tau$ models
  to the ACS/WFC3 $BVizYJH$ photometry of each object pixel individually, versus fitting a simple $\tau$ model to the integrated SED.  Spatially resolved SED modeling results in higher age estimates since variations in SFH on galaxy-wide scales can be accounted for, hence reducing the outshining effect by young stars.
\label{resolv.fig}}
\end {figure}

A second avenue for improvement concerns not so much the wavelength
range, but rather the scales probed with multi-wavelength photometry.
The outshining effect that limits our ability to constrain galaxy ages
results from an older, underlying stellar population whose existence
is not captured by the best-fit SFH because the SED is dominated by
the younger and less obscured part of the stellar population.  If such
multiple components of the galaxy's SFH are spatially disjoint, as
might be expected, we may
bypass or at least reduce the outshining effect by means of resolved
SED modeling.  The advent of the WFC3 camera onboard HST, in concert
with the long-time workhorse ACS, opens a window on such analyses.
Wuyts et al. (in prep) present in detail the implications of resolved
stellar populations at high redshift, but for our present purpose we
focus solely on the estimate of SFR-weighted ages (i.e., the age of
the bulk of the stars).  Here, we use the WFC3 $Y$, $J$, and $H$ data
of the Early Release Science (ERS) program as testbed, in combination
with ACS $B$, $V$, $i$, $z$ data in GOODS-South matched to the same
resolution (0.16", sampled by a 0.06"/pix drizzled pixelscale).  For a
sample of 282 galaxies with $K < 24.3$ and $1.5 < z < 2.5$, we perform
a default SED modeling analysis on the summed BVizYJH photometry of
all galaxy pixels with $S/N_H > 3$.  We ignore any aperture
corrections, which may well be important for estimates of, e.g., total
stellar mass if a substantial fraction of the galaxy mass resides in
the outskirts where pixels have $S/N_H < 3$.  However, it is
irrelevant for our present discussion on galaxy age and how its
estimate depends on an integrated versus resolved approach, because in
both cases we work with the same set of pixels (namely those with
$S/N_H > 3$).  Next, we run the SED modeling procedure on a
pixel-by-pixel basis, yielding a best-fit mass, age, $\tau$, and $A_V$
for each object pixel individually.  Our approach is similar in spirit
to the recent study by Zibetti, Charlot \& Rix 2009 who spatially
resolve the stellar populations in nearby galaxies (obviously to
scales much smaller than probed in our high-redshift sample).  We
confirm that the sum of all best-fit templates from the pixel-by-pixel
analysis provides an excellent match to the integrated galaxy SED.  In
Figure\ \ref{resolv.fig}, we compare the SFR-weighted age obtained
from pixel-by-pixel SED modeling to that preferred by fitting one
$\tau$ model to the integrated galaxy photometry.  Here, the
SFR-weighted age inferred from resolved SED modeling is computed
following equation\ \ref{agew.eq}, but the effective SFH over which one
integrates is now not a simple $\tau$ model, but the superposition of
$\tau$ models, each with their own normalization, $\tau$ and age, for
each of the object pixels.  We observe a clear
trend of increased inferred stellar ages, ranging from 0 to 2 orders
of magnitude, when adopting a resolved SED modeling approach.  The
offsets are small for galaxies where integrated SED modeling indicates
old stellar populations and low SFRs, specific SFRs, and visual
extinctions.  For systems where younger ages, and higher SFRs, SFR/M,
and $A_V$ are inferred from integrated SED modeling, the age
correction factors exhibit a large scatter.  Formally, the offsets
correlate most strongly with galaxy age.  We derive a mean correction
\begin {equation}
\log age_{w, resolved} = 6.86 - 1.75 \log age_{w, integrated}
%^{+ 1.91 - 1.20 \log age_{w, integrated}}
%_{- 1.95 + 0.20 \log age{_{w, integrated}}} .
\end {equation}
with a scatter increasing from 0.1 dex for the systems with the oldest
$age_{w, integrated}$ to 0.4 dex for those with the
youngest $age_{w, integrated}$.
%%%%%%%
% FIG 14
%%%%%%%

This behavior may partially be driven by the condition that the time
since the onset of star formation must be less than the age of the
universe at the epoch of observation (galaxies with old $age_{w,
  integrated}$ already lie close to this upper bound).  We do find only a very weak
correlation with angular size (i.e., the number of independent
resolution elements in which the galaxies are resolved), a property
that itself correlates, at a given mass, with star formation activity
(Franx et al. 2008; Toft et al. 2009).

In order to rule out any systematic biases by fitting to the low S/N
SEDs of individual pixels, we applied the same integrated and resolved
SED modeling procedure to a set of 500 mock galaxies with a uniform
stellar population and representative stellar masses, sizes, and
$H$-band pixel flux distributions.  Our toy galaxies vary in e-folding
time between 30 Myr and 3 Gyr, in the time since the onset of star
formation between 100 Myr and the age of the universe, and were
attenuated by a uniform visual extinction of $A_V = 1$.  We computed
their broad-band SEDs as observed at $z=2$, and disturbed the pixel
fluxes appropriately according to the pixel-to-pixel rms of the real
data.  We confirm that both the integrated and resolved SED modeling
methods correctly recover the input SFR-weighted ages, to within a few
percent.  The central 68th percentile interval of the $\Delta \log
age_w$ distribution broadens by 35\% for resolved compared to
integrated SED modeling, but no systematic bias is introduced by the
method.  Based on this sanity check, we can state with confidence that
the systematic offsets observed for the ERS2 galaxies are due to real
spatial variations in their stellar populations.

Given the limited resolution, we can only account for differences in
SFH across the largest galaxy-wide scales.  The outshining may well
remain present on scales below the resolution of the WFC3 imaging
(0.16" FWHM, corresponding to 1.3 kpc at $z=2$, sampled by drizzled
pixels of 0.06").  We note that the effective visual
attenuation $A_{V, resolved}$, computed as
\begin {eqnarray}
A_{V, resolved} = 2.5 \log \left ( \sum_{i=1}^{Npix} L_{V,intrinsic,i}
\right ) \nonumber \\
 - 2.5 \log \left ( \sum_{i=1}^{Npix} L_{V,attenuated,i}
\right ) ,
\end {eqnarray}
also increases when performing resolved SED modeling, by $0.2 ^
{+0.6}_{-0.2}$ mag.  This technique may therefore help to alleviate
both the outshining effect and the saturation of reddening by patchy
dust obscuration (see Section\ \ref{SEDvsUVIR.sec}).  Obviously, its
use is limited to special data sets rich in multi-wavelength
high-resolution imaging.  Finally, despite the strong observed trends,
we caution that SED modeling on an individual pixel basis is still
subject to the same age-dust degeneracy that is well-known from
integrated SED modeling, particularly when based on the relatively
narrow observed $B$-to-$H$ wavelength baseline.

\section {Summary}
\label{discussion.sec}

%%%%%%%
% FIG 15
%%%%%%%
\begin {figure}[t]
\centering
%\epsscale{1.15}
\plotone{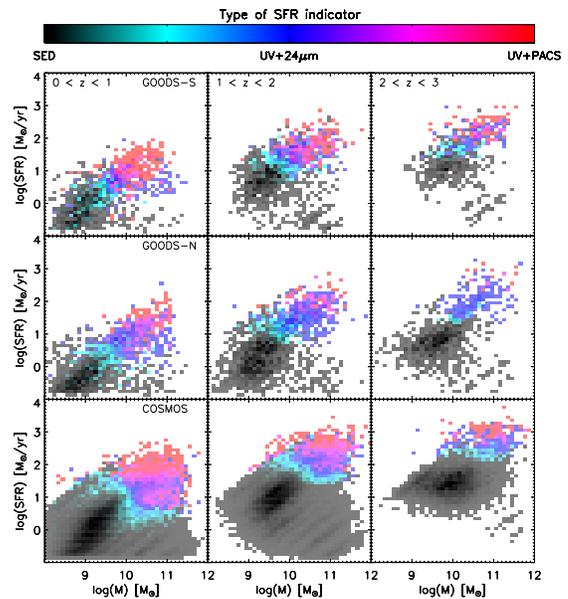} %{figeps/SFRtype.eps}
\caption{
SFR versus mass diagram in the GOODS-South, GOODS-North, and COSMOS
field.  The color-coding illustrates the SFR indicator used, from SED
modeling for all galaxies in a SFR-mass bin ({\it black}) to
$SFR_{UV+PACS}$ for all galaxies in a SFR-mass bin ({\it red}).  Blue
regions contain predominantly galaxies with $SFR_{UV+24\mu m}$ as
primary SFR indicator.  A combination of 
SFR indicators, as cross-calibrated in this paper, is necessary to fully probe the main sequence of star formation, particularly in wider fields with shallower PACS and 
MIPS depths.
\label{SFRtype.fig}}
\end {figure}

We compared SFR indicators out to redshift $z = 3$ relying on PACS
far-infrared, MIPS 24 $\mu$m, and UV emission, $U$-to-8 $\mu$m SED
modeling, as well as H$\alpha$ spectroscopy.  Figure\
\ref{SFRtype.fig} illustrates that, in order to study the entire
galaxy population (above a certain mass) out to high redshift, a
combination of self-consistent SFR indicators is required.  This
applies even more so to studies that exploit larger areas that often
received only shallower PACS and/or MIPS imaging.  This study aims at
providing such a series of cross-calibrated recipes.  In Wuyts et
al. (2011b), we exploit this continuity across SFR indicators in
concert with high-resolution HST imaging to address the relation
between the level and mode of star formation, and its mass dependence.

Using the deeper PEP data in GOODS-South, we confirm previous reports
by Nordon et al. (2010) and Elbaz et al. (2010) that
luminosity-dependent conversions from observed 24 $\mu$m to $L_{IR}$
based on the CE01 and DH02 template libraries lead to overestimated
$L_{IR}$ (and hence SFRs) at the highest SFRs ($SFR > 100\
M_{\sun}/yr$) and redshifts ($z > 1.5$).  This trend is also
consistent with early findings by Papovich et al. (2007) on the basis
of stacking of MIPS 70 $\mu$m and 160 $\mu$m data.  Using the
luminosity-independent conversion by W08, which is based on a template
with enhanced PAH emission relative to local ULIRGs, gives $L_{IR}$
based on 24 $\mu$m that are in the median consistent with PACS
photometry, although with a 0.25 dex scatter.

SED modeling recipes can be tuned to reproduce $SFR_{UV+IR}$ for
systems with low to intermediate SFRs (e.g., by adjusting the minimum
allowed e-folding time for exponentially declining SFHs).  The
relatively long (at least several 100 Myr) e-folding times required
for SED modeling with BC03 to match $SFR_{UV+IR}$ implies that most
star formation happened in a relatively stable mode, varying slowly on
timescales that correspond to of order $\sim 50$ dynamical times.

Galaxies at the high SFR end ($SFR \gtrsim 100\ M_{\sun}/yr$) tend to
be dusty, and are increasingly common toward high redshifts.  Simple
dust correction methods assuming a uniform foreground screen fail to
recover their total amount of star formation.  Due to the patchiness
of the dust distribution in these objects, reddening saturates as a
tracer of extinction.  This bias applies to SED modeling as well as
methods based on single colors, that are designed for a specific
(high) redshift range (Daddi et al. 2007a).  At the low SFR end, we
find a hint of steeper, more SMC-like extinction curves than the
Calzetti et al. (2000) law.  Such a trend could occur if a lower gas
density in those systems lead to a size distribution of the dust
grains mixed with the gas that is biased to small grains.

SFRs based on H$\alpha$ luminosities of star-forming galaxies at $z
\sim 2$ show a good correspondence to $SFR_{UV+IR}$ and $SFR_{SED}$,
provided we account for extra attenuation towards HII regions compared
to the stellar continuum (see also F\"{o}rster Schreiber et al. 2009).
The locally calibrated scaling $A_{V, neb} = A_{V, SED} / 0.44$ holds
out to $z \approx 2.6$.  Since the SINS H$\alpha$ measurements were
obtained in integral field mode, our conclusion is not subject to
uncertainties in slit correction factors.

Next, we fixed the SFR in SED modeling to $SFR_{UV+IR}$, avoiding
biases inherent to dust correction methods and reducing the degrees of
freedom by one.  The SFHs inferred from SED modeling are tested for
consistency with a no-merger galaxy continuity equation.  Briefly, we
compared the observed SFR-M relation and mass function at a range of
redshifts to backtraced analogs based on evolving the observed galaxy
population at lower redshifts back in time.  We find the locus of the
main sequence of star formation to be well reproduced by the
backtraced model.  However, number densities of intermediate- to
low-mass galaxies are underestimated.  This discrepancy is observed
both for simple $\tau$ models and delayed $\tau$ models.  Accounting
for mergers is unlikely to resolve the discrepancy without removing
the agreement at the high-mass end.  A better agreement is only
obtained when forcing early formation redshifts.

Independent evidence from resolved SED modeling also implies that
stellar population modeling of integrated photometry leads to age
underestimates.  Resolving regions that underwent different SFHs helps
to reduce the effect of outshining of old stellar populations by the
youngest generation of stars.  Alternatively, higher $S/N$ data
extending to shorter rest-wavelengths $\lambda_{rest} < 0.2 \mu$m
would provide more leverage on the stellar age.

\vspace{0.2in}
The authors acknowledge Mariska Kriek for use of the FAST stellar
population fitting code.  S. W. wishes to thank Roderik Overzier for
stimulating discussions.  The Millennium Simulation databases used in
this paper and the web application providing online access to them
were constructed as part of the activities of the German Astrophysical
Virtual Observatory.

%\begin {appendix}
\vspace{0.2in}
\begin{center} APPENDIX \end{center}
One concern in exploiting 24 $\mu$m emission as SFR indicator for
high-redshift galaxies is a possible contribution from hot dust heated
by an Active Galactic Nucleus (AGN) rather than young stars.  In order
to investigate the potential bias induced by AGN in our sample, we
mark all galaxies associated with an X-ray source brighter than $L_{X}
> 10^{42}\ erg\ s^{-1}$ in the rest-frame 2 - 10 keV band with a black
open circle in Figure\ \ref{Lir.fig}.  Here, the X-ray luminosities were calculated based on
the galaxy's redshift, and the full band flux and spectral index from
the Luo et al. (2008) 2 Ms Chandra catalog, which we cross-correlated
with the FIREWORKS catalog using a search radius of 1''.  We note
that, owing to the deeper PEP observations in GOODS-South, the
fraction of X-ray AGN with a significant ($> 3\sigma$) PACS detection
is larger ($\sim 35$\%) than the 20\% found in GOODS-North by Shao et
al. (2010).  However, among the total sample of PACS-detected sources,
X-ray AGN form a minority ($< 10$\%).

While the X-ray selected AGN typically have residuals $\nu f_{\nu} / \nu f_{\nu,
  template} < 1$ in Figure\ \ref{Lir.fig}b (for a given IR luminosity, they are brighter at 24
$\mu$m than the overall PACS-detected galaxy population by a median
factor 1.4), excluding them changes the median binned values (filled
black symbols in Figure\ \ref{Lir.fig}b) by less than 8\% only.  We
conclude that, whereas contribution by AGN can significantly affect
the mid- to far-IR SEDs of individual galaxies, their relative number
is sufficiently small that our above result for the ensemble of
PACS-detected galaxies out to $z = 3$ is robust against such a bias.
We do however caution that this assessment of AGN contribution relies
on the assumption that all AGN responsible for hot dust emission have
$L_X > 10^{42}\ erg\ s^{-1}$.  If a substantial population of Compton
thick AGN, undetected at X-ray wavelengths, exists (Daddi et
al. 2007b), other indicators of AGN activity such as a power-law shape
of the SED in the rest-frame 1 - 10 $\mu$m regime (see, e.g., Park et
al. 2010) have to be explored to properly infer SFRs from dust
re-emission.  Exploiting the combination of PACS and Chandra data in
GOODS-North, Shao et al. (2010) find that the far-infrared luminosity
of high-redshift AGN shows little dependence on the X-ray obscuring
column and the AGN luminosity.  Finally, we note that ultra-deep IRS spectroscopy of $z
\sim 1$ LIRGs and $z \sim 2$ ULIRGs (Fadda et al. 2010) found that
both categories are starburst dominated.  Based on the strong PAH
emission in their IRS spectra, Fadda et al. (2010) conclude that the
compton thick AGN contribution to the bolometric luminosity of $z \sim
2$ ULIRGs is substantially lower than previously inferred by Daddi et
al. (2007b).  Nordon et al. (2011) also argue for an enhanced presence
of PAHs in star-forming $z \sim 2$ galaxies, rather than systematic
boosting of 24 $\mu$m fluxes by AGN.
\label{A.app}
%\end {appendix}

% References
\begin{references}
{\small
%\reference{} Baker, A. J., Tacconi, L. J., Genzel, R., Lehnert, M. D.,\& Lutz, D. 2004, ApJ, 604, 125
\reference{} Bell, E. F., McIntosh, D. H., Katz, N.,\& Weinberg, M. D. 2003, ApJS, 133149, 289
\reference{} Bell, E. F., et al. 2005, ApJ, 625, 23
\reference{} Berta, S., et al. 2010, A\&A, 518, 30
\reference{} Borch, A., et al. 2006, A\&A, 453, 869
\reference{} Bouchet, P., Lequeux, J., Maurice, E., Pr\'{e}vot, L.,\& Pr\'{e}vot-Burnichon, M. L. 1985, A\&A, 149, 330
\reference{} Bouwens, R. J., Illingworth, G. D., Franx, M.,\& Ford, H. 2007, ApJ, 670, 928
\reference{} Bouwens, R. J., et al. 2010, in prep (arXiv1006.4360)
\reference{} Brammer, G. B., van Dokkum, P. G.,\& Coppi, P. 2008, ApJ, 686, 1503
\reference{} Bruzual, G,\& Charlot, S. 2003, MNRAS, 344, 1000
\reference{} Bundy, K., et al. 2006, ApJ, 651, 120
\reference{} Calzetti, D., Kinney, A. L.,\& Storchi-Bergmann, T. 1994, ApJ, 429, 582
\reference{} Calzetti, D., Armus, L., Bohlin, R. C., Kinney, A. L., Koornneef, J.\& Storchi-Bergmann, T. 2000, ApJ, 533, 682
\reference{} Cassisi, S., Castellani, M.,\& Castellani, V. 1997, A\&A,
317, 108
\reference{} Chabrier, G. 2003, PASP, 115, 763
\reference{} Chary, R.,\& Elbaz, D. 2001, ApJ, 556, 562
\reference{} Cid-Fernandes, R., Mateus, A., Sodr\'{e}, L. Jr., Stasinska, G.,\& Gomes, J. M. 2005, MNRAS, 358, 363
\reference{} Cole, S., et al. 2001, MNRAS, 326, 255
\reference{} Cowie, L. L.,\& Barger, A. J. 2008, ApJ, 686, 72
\reference{} Daddi, E., Cimatti, A., Renzini, A., Fontana, A., Mignoli, M., Pozzetti, L., Tozzi, P.,\& Zamorani, G. 2004, ApJ, 617, 746
\reference{} Daddi, E., et al. 2007a, ApJ, 670 156
\reference{} Daddi, E., et al. 2007b, ApJ, 670, 173
\reference{} Dale, D. A.,\& Helou, G. 2002, ApJ, 576, 159
\reference{} Damen, M., F\"{o}rster Schreiber, N. M., Franx, M., Labb\'{e}, I., Toft, S., van Dokkum, P. G.,\& Wuyts, S. 2009, ApJ, 705, 617
\reference{} Dav\'{e}, R. 2008, MNRAS, 385, 147
\reference{} De Lucia, G.,\& Blaizot, J. 2007, MNRAS, 375, 2
\reference{} Dickinson, M., Papovich, C., Ferguson, H. C.,\& Budav\'{a}ri, T. 2003, ApJ, 587, 25
\reference{} Di Matteo, T., Springel, V.,\& Hernquist, L. 2005, Nature, 433, 604
\reference{} Drory, N., Bender, R., Feulner, G., Hopp, U., Maraston, C., Snigula, J.,\& Hill, G. J. 2004, ApJ, 608, 742
\reference{} Drory, N., Salvato, M., Gabasch, A., Bender, R., Hopp, U., Feulner, G.,\& Pannella, M. 2005, ApJ, 619, L111
\reference{} Dunne, L., Ivison, R. J., Maddox, S., et al. 2009, MNRAS, 394, 3
\reference{} Elbaz, D., et al. 2007, A\&A, 468, 33
\reference{} Elbaz, D., et al. 2010, A\&A, 518, 29
\reference{} Elmegreen, B. 1997, Rev. Mex. Astron. Astrofis., 6, 165
\reference{} Erb, D. K., Steidel, C. C., Shapley, A. E., Pettini, M., Reddy, N. A.,\& Adelberger, K. L. 2006, ApJ, 647, 128
\reference{} Fadda, D., et al. 2010, ApJ, 719, 425
\reference{} Elsner, F., Feulner, G.,\& Hopp, U. 2008, A\&A, 477, 503
\reference{} Fagotto, F., Bressan, A., Bertelli, G.,\& Chiosi, C. 1994, A\&AS, 104, 365
\reference{} Finlator, K., Dav\'{e}, R.,\& Oppenheimer, B. D. 2007, MNRAS, 376, 1861
\reference{} Fontana, A., et al. 2003, A\&A, 594, L9
\reference{} Fontana, A., et al. 2004, A\&A, 424, 23
\reference{} Fontana, A., et al. 2006, A\&A, 459, 745
\reference{} Fontana, A., et al. 2009, A\&A, 501, 15
\reference{} F\"{o}rster Schreiber, N. M., Genzel, R., Lutz, D., Kunze, D.,\& Sternberg, A. 2001, ApJ, 552, 544
\reference{} F\"{o}rster Schreiber, N. M., et al. 2004, ApJ, 616, 40
\reference{} F\"{o}rster Schreiber, N. M., et al. 2006, ApJ, 645, 1062
\reference{} F\"{o}rster Schreiber, N. M., et al. 2009, ApJ, 706, 1364
\reference{} Franx, M., van Dokkum, P. G., F\"{o}rster Schreiber, N. M., Wuyts, S., Labb\'{e}, I.,\& Toft, S. 2008, ApJ, 688, 770
\reference{} Genzel, R., et al. 2006, Nature, 442, 786
\reference{} Genzel, R., et al. 2008, ApJ, 687, 59
\reference{} Genzel, R., et al. 2010, MNRAS, 407, 2091
\reference{} Giavalisco, M., et al. 2004, ApJ, 600, L103
\reference{} Goldader, J. D., Meurer, G., Heckman, T. M., Seibert, M., Sanders, D. B., Calzetti, D.,\& Steidel, C. C. 2002, ApJ, 568, 651
\reference{} Hopkins, A. M.,\& Beacom, J. F. 2006, ApJ, 651, 142
\reference{} Kajisawa, M., Ichikawa, T., Yamada, T., Uchimoto, Y. K., Yoshikawa, T., Akiyama, M.,\& Onodera, M. 2010, ApJ, 723, 129
\reference{} Kennicutt, R. C. 1998, ARA\&A, 36, 189
\reference{} Kriek, M., et al. 2006, ApJ, 649, 71
\reference{} Kriek, M., van Dokkum, P. G., Labb\'{e}, I., Franx, M., Illingworth, G. D., Marchesini, D.,\& Quadri, R. F. 2009, ApJ, 700, 221
\reference{} Labb\'{e}, I., et al. 2005, ApJ, 624, 81
\reference{} Lee, S.-K., Ferguson, H. C., Somerville, R. S., Wiklind, T.,\& Giavalisco, M. 2010, ApJ, 725, 1644
\reference{} Lilly, S. J., Le F\`{e}vre, O., Hammer, F.,\& Crampton, D. 1996, ApJ, 460, L1
\reference{} Luo, B., et al. 2008, ApJS, 179, 19
\reference{} Madau, P., Ferguson, H. C., Dickinson, M. E., Giavalisco, M., Steidel, C. C.,\& Fruchter, A. 1996, MNRAS, 283, 1388
\reference{} Magdis, G. E., et al. 2010, MNRAS, 409, 22
\reference{} Magnelli, B., Elbaz, D., Chary, R. R., Dickinson, M., Le Borgne, D., Frayer, D. T.,\& Willmer, C. N. A. 2009, A\&A, 496, 57
\reference{} Maiolino, R.,\& Natta, A. 2002, Ap\&SS, 281, 233
\reference{} Maraston, C. 2005, MNRAS, 362, 799
\reference{} Maraston, C., Daddi, E., Renzini, A., Cimatti, A., Dickinson, M., Papovich, C., Pasquali, A.,\& Pirzkal, N. 2006, ApJ, 652, 85
\reference{} Maraston, C., Pforr, J., Renzini, A., Daddi, E., Dickinson, M., Cimatti, A.,\& Tonini, C. 2010, MNRAS, 407, 830
\reference{} Marchesini, D., van Dokkum, P. G., F\"{o}rster Schreiber, N. M., Franx, M., Labb\'{e}, I.,\& Wuyts, S. 2009, ApJ, 701, 1765
\reference{} Meurer, G. R., Heckman, T. M.,\& Calzetti, D. 1999, ApJ, 521, 64
\reference{} Muzzin, A., van Dokkum, P. G., Kriek, M., Labb\'{e}, I., Cury, I., Marchesini, D.,\& Franx, M. 2010, ApJ, 725, 742
\reference{} Noeske, K. G., et al. 2007, ApJ, 660, 43
\reference{} Nordon, R., et al. 2010, A\&A, 518, 24
\reference{} Onodera, M., Arimoto, N., Daddi, E., Renzini, A., Kong, X., Cimatti, A., Broadhurst, T.,\& Alexander, D. M. 2010, ApJ, 715, 385
\reference{} Pannella, M. Carilli, C. L., Daddi, E., et al. 2009, ApJ, 698, L116
\reference{} Papovich, C., et al. 2007, ApJ, 668, 45
\reference{} Papovich, C., Finkelstein, S. L., Ferguson, H. C., Lotz, J. M,\& Giavalisco, M. 2011, MNRAS, 412, 1123
\reference{} Park, S. Q., et al. 2010, ApJ, 717, 1181
\reference{} Peng, C. Y., Ho, L. C., Impey, C. D.,\& Rix, H.-W. 2010, AJ, 139, 2097
\reference{} P\'{e}rez-Gonz\'{a}lez, P. G., et al. 2008, ApJ, 675, 261
\reference{} Poggianti, B. M., Bressan, A.,\& Franceschini, A. 2001, ApJ, 550, 195
\reference{} Poglitsch, A., et al. 2010, A\&A, 518, 2
\reference{} Pozzetti, L., et al. 2007, A\&A, 474, 443
\reference{} Pr\'{e}vot, M. L., Lequeux, J., Pr\'{e}vot, L., Maurice, E.,\& Rocca-Volmerange, B. 1984, A\&A, 132, 389
\reference{} Reddy, N. A.,\& Steidel, C. C. 2009, ApJ, 692, 778
\reference{} Reddy, N. A., Erb, D. K., Pettini, M., Steidel, C. C., Shapley, A. E. 2010, ApJ, 712, 1070
\reference{} Renzini, A. 2009, MNRAS, 398, 58
\reference{} Rettura, A., et al. 2010, ApJ, 709, 512
\reference{} Rodighiero, G., et al. 2010, A\&A, 518, 25
\reference{} Santini, P., et al. 2009, A\&A, 504, 751
\reference{} Schiminovich, D., et al. 2005, ApJ, 619, L47
\reference{} Schmidt, M. 1959, ApJ, 129, 243
\reference{} Shao, L., et al. 2010, A\&A, 518, 26
\reference{} Shapley, A. E., Steidel, C. C., Erb, D. K., Reddy, N. A., Adelberger, K. L., Pettini, M., Barmby, P.,\& Huang, J. 2005, ApJ, 626, 698
\reference{} Silk, J. 1997, ApJ, 481, 703
\reference{} Somerville, R. S., Lee, K., Ferguson, H. C., Gardner, J. P., Moustakas, L. A.,\& Giavalisco, M. 2004, ApJ, 600, 171
\reference{} Springel, V., et al. 2005, Nature, 435, 629
\reference{} Steidel, C. C., Adelberger, K. L., Giavalisco, M., Dickinson, M.,\& Pettini, M. 1999, ApJ, 519, 1
\reference{} Symeonidis, M., Page, M. J.,\& Seymour, N. 2011, MNRAS, 411, 983
\reference{} Toft, S., Franx, M., van Dokkum, P., F\"{o}rster Schreiber, N. M., Labb\'{e}, I., Wuyts, S.,\& Marchesini, D. 2009, ApJ, 705, 255
\reference{} Vergani, D., et al. 2008, A\&A, 487, 89
\reference{} Wolf, C., Meisenheimer, K., Rix, H.-W., Borch, A., Dye, S.,\& Kleinheinrich, M. 2003, A\&A, 401, 73
\reference{} Wuyts, S., et al. 2007, ApJ, 655, 51
\reference{} Wuyts, S., Labb\'{e}, I., F\"{o}rster Schreiber, N. M., Franx, M., Rudnick, G., Brammer, G. B.,\& van Dokkum, P. G. 2008, ApJ, 682, 985
\reference{} Wuyts, S., Franx, M., Cox, T. J., Hernquist, L., Hopkins, P. F., Robertson, B. E.,\& van Dokkum, P. G. 2009a, ApJ, 696, 348
\reference{} Wuyts, S., et al. 2009b, ApJ, 700, 799
\reference{} Yoshikawa, T., et al. 2010, ApJ, 718, 112

}
\end {references}

% Tables

% Figures
% FIG 1

% FIG 2

% FIG 3

% FIG 4

% FIG 5

% FIG 6

% FIG 7

% FIG 8

% FIG 9

% FIG 10

% FIG 11

% FIG 12

% FIG 13

% FIG 14

% FIG 15

\end {document}